\documentclass[fleqn,usenatbib]{mnras}

\usepackage{newtxtext,newtxmath}

\usepackage[T1]{fontenc}

\DeclareRobustCommand{\VAN}[3]{#2}
\let\VANthebibliography\thebibliography
\def\thebibliography{\DeclareRobustCommand{\VAN}[3]{##3}\VANthebibliography}

\usepackage{graphicx}	%
\usepackage{amsmath}	%
\usepackage{xcolor}	%

\graphicspath{{figures/}}

\newcommand{\quokka}{\textsc{quokka}}
\newcommand{\msun}{\mathrm{M}_{\odot}}
\newcommand{\zsol}{Z_{\odot}}
\newcommand{\siggas}{\Sigma_{\rm gas}}
\newcommand{\Sigmas}{\Sigma_{*}}
\newcommand{\him}{{\textsc{H}\,\textsc{i}}}
\newcommand{\kmps}{\mathrm{km\,s^{-1}}}
\newcommand{\accunit}{\mathrm{km\,s^{-1}\,Myr^{-1}}}

\defcitealias{QEDI}{QED I}
\defcitealias{QEDIII}{QED III}
\defcitealias{QEDV}{QED V}

\title[Neutral hydrogen in supernova-driven outflows]{QED VI: Neutral hydrogen gas in supernova-driven galactic outflows}

\author[C.-C. He et al.]{
Chong-Chong He,$^{1}$\thanks{E-mail: chongchong.he@anu.edu.au}
Mark R. Krumholz,$^{1}$
Benjamin D. Wibking$^{2}$
and Aditi Vijayan$^{3}$
\\
$^{1}$Research School of Astronomy and Astrophysics, Australian National University, 233 Mount Stromlo Road, Stromlo, ACT 2612, Australia\\
$^{2}$Carnegie Observatories, 813 Santa Barbara St, Pasadena, CA 91101, USA\\
$^{3}$School of Earth and Space Exploration, Arizona State University, Tempe, AZ 85281, USA
}

\date{Accepted XXX. Received YYY; in original form ZZZ}

\pubyear{\the\year{}}

\begin{document}
\label{firstpage}
\pagerange{\pageref{firstpage}--\pageref{lastpage}}
\maketitle

\begin{abstract}
Winds that eject mass from galactic discs regulate the growth of galaxies, and observations show that, at least within a few kpc of the disc, most of the mass flux is carried by neutral rather than ionized material. However, it has proven difficult for simulations of supernova-driven winds to reproduce this observation, leading to speculations that the bulk of the mass must be driven by radiation pressure or cosmic rays rather than supernovae. Here we report simulations of supernova-driven winds using the GPU-accelerated code \quokka\ that reach a resolution of $\Delta x = 1$~pc, significantly higher resolution and simulation volume than previous work. At this resolution the winds are dominated by neutral mass, with cloud sizes, masses and column densities broadly similar to those of the H\,\textsc{i} clouds in the Milky Way's nuclear wind. The neutral material is not ballistic, but is instead driven by momentum transfer associated with drag and condensation of hot gas onto the neutral phase. By contrast, when we coarsen the resolution to values comparable to those used in earlier work, the neutral outflow weakens at $\Delta x = 2$~pc and disappears at $\Delta x = 4$~pc. We argue that this difference arises partly because lower-resolution simulations do not resolve, with a few cells, the $3$--$6$~pc thick cooling layers in which hot gas condenses onto neutral clouds.
\end{abstract}


\section{Introduction}
\label{sec:intro}

Galactic winds that remove mass and metals from star-forming discs play a decisive role in galaxy evolution, and in gas-rich regions much of the mass flux appears to be carried by the neutral and molecular phases \citep{Leroy15b, Martini18a, Krieger19a, Veilleux20a, Yuan23a}.
The neutral material in these outflows extends to distances of at least several kpc above the disc. Observations of the Milky Way's own nuclear wind, where they can achieve the highest resolution, reveal that the neutral material consists of a mist of clouds with typical radii of $\sim 10-30$ pc and masses from $10$ to $10^5\,\msun$ \citep{McClureGriffiths2013, DiTeodoro2018, Lockman2020, Noon2023}, and that some of the H~\textsc{i} clouds host cold molecular material at their centres \citep{DiTeodoro2020, Noon2023}. Atomic clouds have been detected up to $2.5$--$3.5$~kpc off the plane, with velocities that rise from $\approx150$--$200$ to $\approx330\,\kmps$ and imply lifetimes of $4$--$10$~Myr \citep{Lockman2020}, and in chemical states that suggest the material must have been entrained from the Galactic plane rather than condensing {\it in situ} \citep{Noon26a}, as in the cooling-flow scenario of \citet{Schneider2018}.

The survival of neutral clouds to such distances and times has proven difficult to explain theoretically: an adiabatic, non-magnetised cold cloud embedded in a hot wind will be disrupted on a cloud-crushing timescale that is far shorter than the inferred cloud lifetimes, and is never accelerated to velocities as high as those seen \citep{Klein1994}; even with radiative cooling, clouds near a starburst are destroyed before they reach $\gtrsim 1$~kpc \citep{SchneiderRobertson2017}. Radiative cooling  \citep{GronkeOh2018}, thermal conduction \citep{Bruggen2016} and magnetic fields \citep{McCourt2015,Cottle2020} may mitigate these effects, but have mostly been demonstrated in idealised ``wind-tunnel'' simulations that follow a single cloud. A few simulations of full galactic winds also show cool gas gaining mass or momentum from the hot phase, through mixing in starburst-driven outflows \citep{Schneider2020, SchneiderMao2024}, mixing-layer growth in clustered-supernova tall boxes \citep{TanFielding2024}, and entrainment of H~\textsc{i} clouds above the Galactic centre \citep{Peschken2026}. Nevertheless, stratified-box simulations, which follow a vertically extended patch of a star-forming disc and are the standard tool for studying how galactic winds are launched, have had great difficulty reproducing the observed abundance of neutral gas in winds in simulations where the winds are driven by supernovae \citep{Walch2015,Girichidis2016, KimTIGRESSpp2026, KimSMAUG2020, Peschken2026}. In such simulations, while neutral material dominates very close to the disc, almost all the gas is ionized by the time it reaches distances $\gtrsim 1$ kpc from the galactic plane. Their typical neutral gas mass loading factors -- defined as the ratio of the wind mass flux to the star formation rate -- at distances of $1-2$ kpc are $\eta_{\him} \lesssim 10^{-4}$ \citep[e.g.,][]{KimSMAUG2020, Peschken2026}. Significant amounts of neutral material are found only in simulations where the outflow is assumed to be driven primarily by non-thermal pressure sources such as cosmic rays \citep[e.g.,][]{Armillotta24a, Thomas25a, Sike25a}.

However, typical resolutions in the stratified-box simulations carried out to date are coarse compared with the observed $\sim 10-30$ pc sizes of neutral clouds. For example the SILCC simulation suite reaches only $\approx4$~pc resolution \citep{Girichidis2016}, while the TIGRESS suite reaches $2$~pc \citep{Kim2017}, but only within $|z| = 2$~kpc from the disc, and only for small disc patches of well under 1 kpc in size. Moreover, due to the numerical difficulty of simulating flows with such extreme density contrasts, many of these simulations resort to relatively diffusive numerical methods, for example piecewise-linear rather than piecewise-parabolic interpolation, further lowering the effective resolution. The highest-resolution simulations of this type carried out to date that include neutral gas are the QED simulation suite \citep[hereafter \citetalias{QEDI}, \citetalias{QEDIII}, and \citetalias{QEDV}, respectively]{QEDI, QEDIII, QEDV}, and there is preliminary evidence from both the QED simulations and other studies \citep{Kim2017, SchneiderRobertson2018, Peschken2026} that the properties of cool material in outflows are sensitive to resolution. However, no study has yet demonstrated convergence in the neutral content of a supernova-driven outflow.

Here we report a series of simulations of a supernova-driven outflow from a GPU-accelerated code \quokka\ \citep{Wibking2022, He2024a, He2024b, He2026} in which we reach resolutions of $1$ pc at distances of $\pm 4$ kpc from a galactic plane.
Our scheme employs piecewise-parabolic reconstruction, which provides a more accurate representation of smooth flows than the piecewise-linear methods commonly used in earlier stratified-box studies, and thus yields sharper shocks and higher effective resolution for advection at fixed grid spacing \citep[e.g.][]{Leidi2024}. These simulations therefore, for the first time, achieve spatial resolutions sufficient to resolve the observed sizes of neutral clouds at their detected distances.
We carry out a resolution study in which we hold the physical model fixed and vary only the grid
spacing.
We show that at our highest resolution our simulations produce a neutral mass-dominated wind whose clouds have sizes, masses and column densities broadly similar to those of the H\,\textsc{i} clouds observed in the Milky Way's nuclear wind, while our lower resolution cases, similar to earlier work, fail to produce a neutral-dominated wind. We analyse the simulations to understand what drives neutral winds and why they have been missed in earlier works.

The paper is organised as follows. Section~\ref{sec:sims} describes the simulations, and 
we present our results in Section~\ref{sec:results}, discuss their implications and limitations in Section~\ref{sec:discussion}, and summarise our conclusions in Section~\ref{sec:conclusions}.

\section{Simulations}
\label{sec:sims}

\subsection{Initial conditions}
\label{sec:ics}

We evolve a stratified patch of a galactic disc, using periodic boundary conditions in the horizontal ($x$ and $y$) directions, and ``diode'' boundary conditions that permit outflow but suppress inflow in the $z$ direction -- see \citetalias{QEDI} for details. The domain is $0.98\times 0.98$~kpc horizontally and spans $|z| \le 3.92$~kpc, or $0.98\times 0.98\times\pm 3.92$~kpc, and we use three cell counts: $1024^2\times8192$, $512^2\times4096$, and $256^2\times2048$, giving spacing of $\Delta x = 0.96$ pc, 1.91 pc, and 3.83 pc, respectively; for brevity we will refer to these cases as $\Delta x = 1$, 2, and 4 pc. For comparison with previous work, the TIGRESS-NCR LGR4 model reaches $\Delta x = 2$~pc in a $512^2\times3072$~pc domain \citep{Kim2023NCR}, so our $1$~pc run resolves roughly nine times the volume at half the cell size.

All runs begin with a gas surface density $\siggas = 50\,\msun\,\mathrm{pc}^{-2}$, with the gas initially uniform in $x$ and $y$ and stratified in $z$ with a vertical profile chosen to be in hydrostatic equilibrium. The gas initially has a uniform temperature $T=10^4$ K, and a metallicity $Z = 1\,\zsol$. In addition to the gravitational potential generated by solving the Poisson equation for the gas and stellar particles (see next section), our simulations also include an ``external'' potential that combines
a stellar disc and a dark-matter halo,
\begin{equation}
\Phi_{*\text{-}\mathrm{dm}} = 2\pi G\Sigma_* z_*
  \left[\left(1 + \frac{z^2}{z_*^2}\right)^{1/2} - 1\right]
  + 2\pi G\rho_\mathrm{dm} R_0^2\ln\left(1 + \frac{z^2}{R_0^2}\right),
\label{eq:phiext}
\end{equation}
where we adopt $\Sigmas = 208$ M$_\odot$ pc$^{-2}$, $\rho_\mathrm{dm} = 2.4\times 10^{-2}$ M$_\odot$ pc$^{-3}$, $z_* = 245$ pc, and $R_0 = 4$ kpc; these parameters are identical to those in the R4 model of \citet{KimSMAUG2020}. These gas and stellar surface densities are comparable to those of moderately starbursting regions such as the Milky Way centre. See \citetalias{QEDI} for further details on the model setup.

We run each simulation for the times indicated in Table~\ref{tab:runs}, writing the full 3D structure on average every $2$, $3$ and $4$~Myr at $\Delta x = 1$, $2$ and $4$~pc, and reduced 2D projections of select quantities at a cadence of 0.5 Myr.

\begin{table}
\centering
\caption{Summary of all simulations, including the total run duration $t_\mathrm{end}$, time-averaged star formation rate $\langle \mathrm{SFR}\rangle$, and time-averaged total, H~\textsc{i}, hot, and WIM mass-loading factors $\eta_\mathrm{total}$, $\eta_{\him}$, $\eta_\mathrm{hot}$, and $\eta_\mathrm{WIM}$, and the atomic mass-loading factor $\eta_{\rm atom}$ (Section~\ref{sec:phases}), computed at the outer edge of the simulation box. All time-averaged quantities, including $\langle \mathrm{SFR}\rangle$, are averaged over the interval from $29$~Myr, about one free-fall time at the box half-height, to $112$~Myr, the last three-dimensional output of the 1~pc run.}
\label{tab:runs}
\setlength{\tabcolsep}{4pt}
\begin{tabular}{cccccccc}
\hline
$\Delta x$ & $t_{\rm end}$ & $\langle\mathrm{SFR}\rangle$ &
  $\eta_{\rm total}$ & $\eta_{\him}$ & $\eta_{\rm hot}$ &
  $\eta_{\rm WIM}$ & $\eta_{\rm atom}$ \\
$[$pc$]$ & [Myr] & [$\msun\,\mathrm{yr}^{-1}$] & & & & & \\
\hline
$1$ & $114$ & $0.178$ & $0.735$ & $0.316$ & $0.247$ & $0.172$ & $0.398$ \\
$2$ & $176$ & $0.277$ & $0.405$ & $0.117$ & $0.176$ & $0.112$ & $0.170$ \\
$4$ & $256$ & $0.364$ & $0.061$ & $0.000$ & $0.061$ & $0.000$ & $0.000$ \\
\hline
\end{tabular}
\end{table}

\subsection{Numerical method}
\label{sec:numerics}

We carry out all simulations with \quokka, a GPU-native, block-structured adaptive-mesh radiation-hydrodynamics code built on AMReX \citep{Wibking2022,He2024a,He2024b,He2026}. We use a limited subset of its capabilities: the radiation-transport module is switched off, and mesh refinement is not used, so each run is a single uniform Cartesian grid. The code solves the Euler equations of hydrodynamics coupled to the Poisson equation for self-gravity (along with an external gravitational potential), optically thin radiative cooling, photoelectric heating and star-particle feedback. The hydrodynamics uses a piecewise-parabolic reconstruction of the primitive variables \citep{Colella1984} with the HLLC Riemann solver, a method-of-lines formulation and second-order strong-stability-preserving Runge--Kutta time-stepping.

Star formation and supernova injection are self-consistent rather than prescribed, and use the ``particle--mesh--particle'' scheme of \citet{He2026}. In cells that satisfy a local Jeans-instability criterion, the code stochastically forms a collisionless star particle at a rate set by the local free-fall efficiency, drawing stellar masses from a Chabrier initial mass function \citep{Chabrier2003}; stars massive enough to explode as supernovae ($>9$ M$_\odot$ following the models of \citealt{Sukhbold16a}) are followed individually, while lower-mass stars are lumped together into a composite particle. The resulting star particles are given an initial velocity distribution that includes a tail to high velocities corresponding to runaway stars (see \citealt{He2026}), and move freely through the domain under the effects of gravity. Once a particle in the range of initial masses predicted by \citeauthor{Sukhbold16a} to explode as supernovae reaches an age that exceeds its progenitor's main-sequence lifetime (as determined from the MIST stellar evolution tracks -- \citealt{Choi16a}) it injects $\Delta E_{\rm SN} = 10^{51}$~erg, $\Delta M_{\rm SN} = 10\,\msun$ and $\Delta M_{Z,\rm SN} = 1\,\msun$ of energy, mass, and metals into its host cell; we treat metals as a passive scalar. Supernova feedback uses a momentum-energy switch, as discussed in \citet{He2026}: if the energy-conserving (Sedov--Taylor) phase of the remnant is resolved, the code deposits the supernova energy, and if it is not, the code deposits the terminal momentum of the remnant instead. At our highest resolution the momentum branch is almost never used, and so in most cases supernova feedback is handled as pure energy deposition. At $\Delta x = 4$~pc the Sedov--Taylor phase is unresolved more often, so the momentum branch is used more often; we discuss the consequences in Section~\ref{sec:resolution}.

While we do not include radiative transport, we do include radiative heating and cooling using the \textsc{Grackle} library \citep{Smith2017}. Specifically, we use the \textsc{Grackle} table \texttt{CloudyData\_UVB=HM2012\_shielded.h5}, computed with \textsc{Cloudy} for the uniform metagalactic ultraviolet background of \citet{Haardt2012} at redshift zero. In this table the background is attenuated by self-shielding: each density--temperature pair is modelled as a slab whose depth is the local Jeans length, capped at $100$~pc \citep{Smith2017, Emerick2019}. We use the table in its fully tabulated equilibrium mode, the equivalent of \textsc{Grackle}'s \texttt{primordial\_chemistry = 0}, in which no species abundances are evolved and the net heating and cooling rate and the temperature are functions of density and internal energy only. The tables include primordial (H and He) cooling and heating, metal-line cooling and heating, and inverse Compton heating by cosmic microwave background photons; as in \citetalias{QEDV}, we hold the metal line cooling table at a fixed metallicity $Z=Z_\odot$ rather than evolving it in response to metal deposition by supernovae. We recompute the table as a function of ($\rho, E_{\rm int}$) rather than the default ($\rho, T$) for computational efficiency, but the underlying physics is identical. We discuss the table in more detail, and address some potential caveats regarding the choice of shielding prescription we implicitly make by using it, in Appendix~\ref{sec:table}.

We make one additional modification to the cooling table, which is that we replace its constant photoelectric heating term with time-dependent photoelectric heating rate that tracks the instantaneous, spatially uniform FUV field produced by the simulated stellar population; this rate enters only the energy equation and does not change the ionization state, which the table sets. The instantaneous photoelectric heating rate per H nucleon depends on star formation history in the simulation box as
\begin{equation}
  \Gamma_{\rm PE}(t) = \Gamma_{\rm PE,0} \frac{\int_0^{t} \psi(t - t') \dot{\Sigma}_*(t')\, dt'}{\int_0^{1~\rm Gyr} \psi(t-t') \dot{\Sigma}_{*,0} \, dt'}
\end{equation}
where $\Gamma_\mathrm{PE,0} = 2.0\times 10^{-26}$ erg s$^{-1}$ H$^{-1}$ is the Solar neighborhood photoelectric heating rate \citep{Koyama2002}, $\dot{\Sigma}_*(t)$ is the star formation rate per unit area in the simulation box as a function of time, $\psi(t)$ is the FUV luminosity per unit mass produced by a stellar population of age $t$, and $\dot{\Sigma}_{*,0} = 2.5 \times 10^{-3}$ M$_\odot$ yr$^{-1}$ kpc$^{-2}$ is the star formation rate in the Solar neighbourhood \citep{Fuchs09a}. We compute $\psi(t)$ by using the \textsc{slug} stellar population synthesis code \citep{da-Silva12a, Krumholz15b} to compute the total luminosity in photons with energies $E_\gamma = 8 - 13.6$ eV (the approximate energy range responsible for photoelectric heating) for a continuously-sampled stellar population as a function of age $t$, using MIST stellar tracks \citep{Choi16a}, the same Chabrier IMF used in the simulations, and \textsc{slug}'s ``starburst99'' spectral synthesis mode \citep{Leitherer99a, Vazquez05a}. Note that the exact choice of 1 Gyr as the upper limit in the integral in the denominator has almost no effect on the value of the integral, since $\psi(t)$ goes to near-zero values for $t \gtrsim 200$ Myr; thus any value greater than a few hundred Myr yields the same result to within a few percent. Because there is no stellar population and thus no photoelectric heating at the start of a simulation, we add an artificial photoelectric heating source for the first $\sim$free-fall time of each simulation in order to maintain a two-phase equilibrium structure while the simulation settles to equilibrium. Specifically, we add a uniform heating rate of $1.3\times10^{-25}$~erg~s$^{-1}$~H$^{-1}$ during the first $16$~Myr of each simulation, which then declines linearly to a negligible floor at $33$~Myr, shortly after the start of our averaging interval.

\subsection{Phase decomposition and outflow rates}
\label{sec:phases}

Our primary concern in this work is with the phase structure of gas in the outflow. In analysing our results we will therefore always split the gas into three phases that partition it exactly: \emph{neutral}, meaning gas with electron abundance $x_e < 0.05$; \emph{hot}, meaning $T > 10^6$~K; and the \emph{warm ionised medium} (WIM), the remainder. We label quantities of the neutral phase with the subscript H\,\textsc{i} (e.g.\ $\eta_{\him}$); like those of the other phases, they refer to the total gas mass, including helium. Our simulations do not follow a non-equilibrium chemical network, so we reconstruct $x_e$ from the tabulated equation of state from \textsc{Grackle} (Section~\ref{sec:numerics}), which gives the temperature as a function of gas density and specific internal energy $\varepsilon$; the mean molecular weight then follows from $T/\mu = (\gamma - 1)\, m_{\rm u}\,\varepsilon / k_{\rm B}$, with $\gamma = 5/3$. For \textsc{Grackle}'s adopted hydrogen mass fraction $X = 10/14$ and helium mass fraction $Y = 4/14$, the electron abundance per H nucleon $x_e \equiv n_e/n_{\rm H}$ is then related to the mean molecular weight as $x_e = (\mu^{-1} - Y/4)/X - 1$, where $x_e$ counts the free electrons from both hydrogen and helium. We deliberately define the neutral phase in terms of the electron abundance rather than simply using a temperature cut because the temperature at which the gas transitions from predominantly ionized to predominantly neutral is density-dependent, ranging from $10^{3.8}$--$10^{4.2}$~K, so using a simple temperature cut would not differentiate cleanly between ionized and neutral material at similar temperatures. For the purposes of comparing to observations in the 21 cm line, our primary source of information on the detailed structure of the neutral material entrained into outflows, the ionization fraction is the important quantity.

For each phase $p$ the mass outflow rate at height $h$ is the net flux through the pair of planes at $z = \pm h$. Formally, we evaluate this as
\begin{equation}
\dot{M}_p(h,t) = \int_{z=+h}\rho_p v_z\,\mathrm{d}A
  - \int_{z=-h}\rho_p v_z\,\mathrm{d}A,
  \label{eq:mass_flux}
\end{equation}
where $\rho_p$ is the gas density for cells classified as belonging to phase $p$ and 0 otherwise, and $v_z$ is the vertical velocity. We evaluate this quantity at the quarter, half, three-quarter and full box half-height, $h = 0.98$, $1.96$, $2.93$ and $3.91$~kpc, which in the text we round to $1$, $2$, $3$ and $4$~kpc for brevity. We define the corresponding mass loading factor of each phase as
\begin{equation}
    \eta_p = \frac{\dot{M}_p}{\langle\mathrm{SFR}\rangle},
\end{equation}
where $\langle\mathrm{SFR}\rangle$ is the time-averaged star formation rate.
Note that our outflow rates are net mass fluxes rather than the outflow-only values reported in some works \citep[e.g.,][who replace $v_z$ with $\max(v_z, 0)$ in the first integral and $\min(v_z, 0)$ in the second]{Rathjen2023}, so comparison must be made with care.

The threshold $x_e < 0.05$ is strict: it assigns to the WIM gas that is partly ionized but still mostly atomic. Observations in the 21~cm line, by contrast, measure the atomic hydrogen in all of the gas, whatever its ionized fraction. We therefore also report the \emph{atomic} mass flux $\dot{M}_{\rm atom}$, defined by equation~(\ref{eq:mass_flux}) with $\rho_p$ replaced by $\rho\,(1 - x_e)$ with $1 - x_e$ clipped to the interval $[0, 1]$, and the corresponding loading factor $\eta_{\rm atom}$. This weights each cell by its atomic fraction instead of assigning it wholly to one phase. Because $x_e$ includes the electrons from helium, $1 - x_e$ slightly underestimates the atomic fraction of hydrogen where helium is ionized. Maps, vertical profiles and cloud properties require a binary classification, so for these we use the neutral phase defined by the threshold; for the mass loading we give both $\eta_{\him}$ and $\eta_{\rm atom}$.

We measure the outflow rates summed over all phases and for the hot phase at 0.5 Myr intervals, and the outflow rate in the neutral phase and the atomic mass flux at the coarser cadence of the 3D outputs, since it is more expensive to compute because $x_e$ must be obtained in post-process rather than on-the-fly as the simulation runs. The WIM rate is the remainder, $\dot{M}_{\rm WIM} = \dot{M}_{\rm total} - \dot{M}_{\him} - \dot{M}_{\rm hot}$, so wherever the three are compared they carry the neutral measurement's cadence.

Throughout this work, all time-averaged quantities -- including the star formation rate used to normalise $\eta_p$ -- are computed over a common interval of $29$--$112$~Myr. The starting time is chosen to exclude an initial transient phase, which lasts for roughly one free-fall time at the box half-height. We find that star formation in the simulation is bursty and episodic once it has begun, with period of $\approx 120$ Myr. We end the averaging interval at $112$~Myr, the last three-dimensional output of the $1$~pc run.

\section{Results}
\label{sec:results}
\begin{figure*}
\centering
\includegraphics[width=\textwidth,height=0.8\textheight,keepaspectratio]{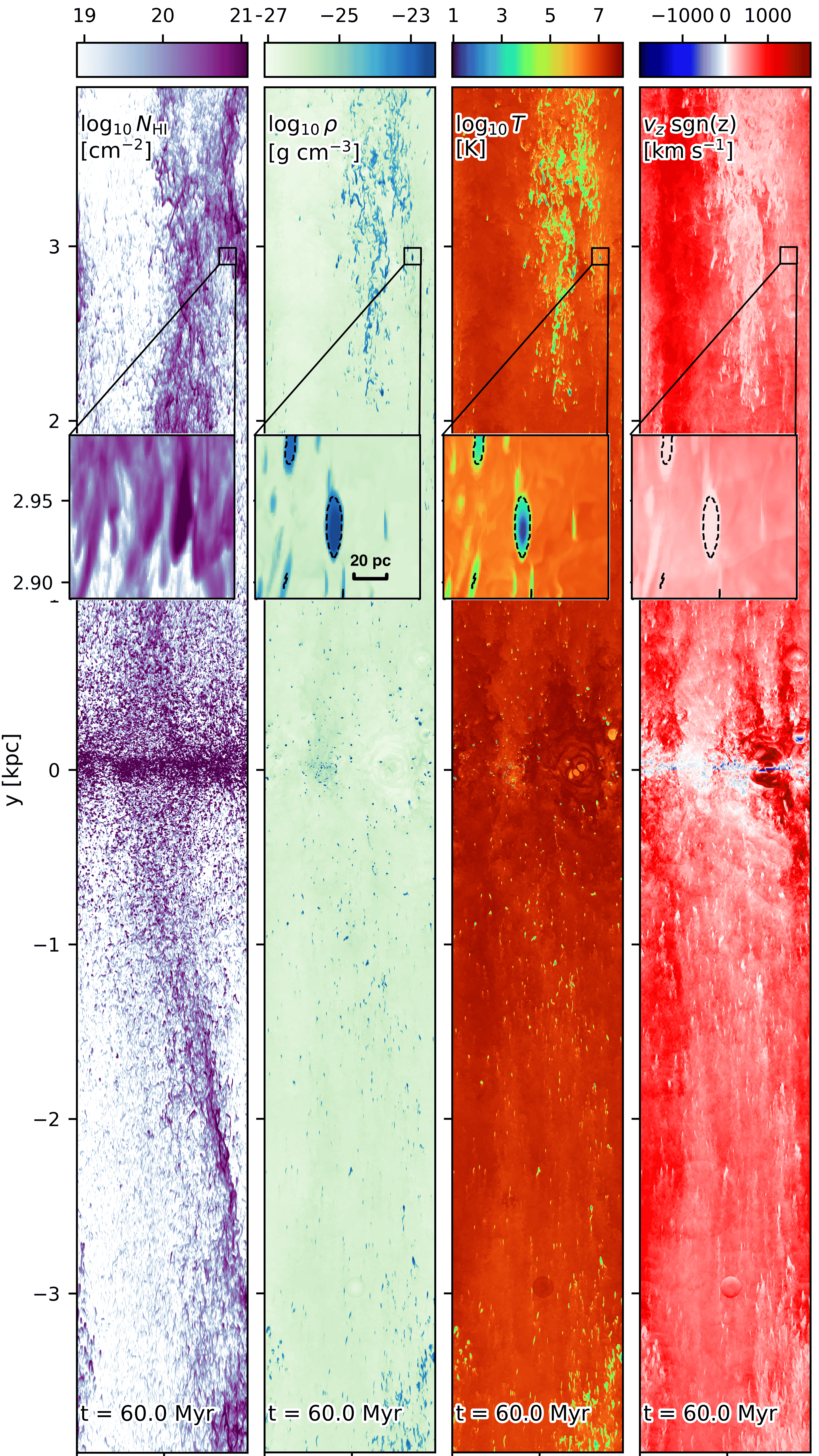}
\caption{Edge-on view of the 1 pc resolution simulation at $t=60$ Myr. Columns show, from left to right, the projected neutral gas column $N_{\him}$, total density $\rho$ in a slice, gas temperature $T$ in a slice, and gas vertical velocity $v_z$ in a slice, coloured so that velocities away from the midplane are red. The vertical axis, labelled $y$ in this figure and in Fig.~\ref{fig:res}, is the height $z$ above the mid-plane. Insets zoom in on one example neutral cloud $\approx 2.9$ kpc above the plane, with dashed black contours marking the locus $x_e = 0.05$ that we use to define the transition from ionized to neutral.}
\label{fig:morph}
\end{figure*}

\begin{table}
\centering
\caption{Neutral mass loading as a function of height and resolution. Here $h$ is the height, $\eta_{\him}$ is the neutral mass loading factor, and $\eta_{\him}/\eta_\mathrm{total}$ is the fraction of the total mass flux carried by neutral material. All quantities are time averages over $29$--$112$~Myr.}
\label{tab:ladder}
\setlength{\tabcolsep}{4pt}
\begin{tabular}{crcccrccc}
\hline
$h$ & & \multicolumn{3}{c}{$\eta_{\him}$} & &
   \multicolumn{3}{c}{$\eta_{\him}/\eta_{\rm total}$} \\
{[kpc]} & $\Delta x =$\!\!\!\! & $4$~pc & $2$~pc & $1$~pc & $\Delta x={}$\!\!\!\! & $4$~pc & $2$~pc & $1$~pc \\
\hline
$0.98$ & & $-0.001$ & $0.118$ & $0.262$ & & $-0.020$ & $0.270$ & $0.349$ \\
$1.96$ & & $0.000$ & $0.110$ & $0.257$ & & $0.004$ & $0.258$ & $0.354$ \\
$2.93$ & & $0.000$ & $0.113$ & $0.306$ & & $-0.001$ & $0.272$ & $0.420$ \\
$3.91$ & & $0.000$ & $0.117$ & $0.316$ & & $0.001$ & $0.288$ & $0.430$ \\
\hline
\end{tabular}
\end{table}

\subsection{Full resolution results}
\label{sec:fullres}

Figure~\ref{fig:morph} shows the flagship $1$~pc simulation at $t = 60$~Myr, close to the time of maximum mass outflow rate. It is clear from the figure that neutral gas reaches the top of the domain $4$~kpc above the disc, and that it is structured as clouds and filaments tens of parsecs in size, embedded in a volume-filling wind moving faster than $500\,\kmps$. Neutral gas dominates the mass of the outflow at every height we measure: at $t = 60$~Myr it holds $69$ per cent of the gas mass above $|z| = 1$~kpc ($64$ per cent averaged over $29$--$112$~Myr), against $\approx 20$ per cent in the warm ionised medium and $10$ per cent in the hot phase. Its share of the time-averaged mass flux is smaller, $35$--$43$ per cent, and rises with height, while the hot phase carries almost half of the flux at $1$~kpc and a third at the box edge, so the neutral phase carries the largest share of the flux only at $|z| \gtrsim 3$~kpc; see Section~\ref{sec:phases} for how we define each phase, and Table~\ref{tab:ladder} for exact numerical values. Crucially, neither the neutral mass loading nor the fraction of outflowing mass carried in the neutral phase decay with height: $\eta_{\him} \approx 0.3$ and $\eta_{\him}/\eta_\mathrm{total} \approx 0.4$ at all heights from $1 - 4$ kpc. Weighting each cell by its atomic fraction instead (Section~\ref{sec:phases}) gives $\eta_{\rm atom} \approx 0.3$--$0.4$, so that atomic gas carries $41$--$54$ per cent of the mass flux at these heights.

\begin{table*}
\centering
\caption{Structural properties of neutral clouds, defined as connected sets of cells with electron fraction $x_e < 0.05$, in three $600$~pc-thick slabs centred at the $z$ value, from five snapshots spanning $49$--$70.6$~Myr. \textit{Population} quantities characterise the cloud population as a whole, \textit{Count-weighted} and \textit{Mass-weighted} properties give the median and, in brackets, the 25th to 75th percentile range of each cloud property, weighted by number and by mass (mass-weighted percentiles are computed in each snapshot, and the median over the five snapshots is given), and \textit{Survival criterion} lists quantities used to evaluate the \citet{GronkeOh2018} cloud survival criterion, all computed from the number-weighted medians. See Sections~\ref{sec:fullres} and \ref{sec:survival} for definitions.
}
\label{tab:clouds}
\begin{tabular}{lccc}
\hline
 & \multicolumn{3}{c}{$z$ [kpc]} \\
 & $1.5$ & $2.5$ & $3.5$ \\
\hline
\multicolumn{4}{l}{\emph{Population}} \\
$n_{\rm cl}$ [kpc$^{-3}$] & $1541$ & $1053$ & $965$ \\
$f_M(R_\mathrm{eff}\!\ge\!30\,\mathrm{pc})$ & $52\%$ & $70\%$ & $73\%$ \\
\hline
\multicolumn{4}{l}{\emph{Count-weighted}} \\
$R_{\rm eff}$ [pc] & $6.0\,[5.3\text{--}7.4]$ & $6.2\,[5.3\text{--}7.7]$ & $6.2\,[5.3\text{--}7.9]$ \\
$M$ [$\msun$] & $96\,[54\text{--}215]$ & $58\,[34\text{--}134]$ & $45\,[26\text{--}104]$ \\
$n_{\rm H}$ [cm$^{-3}$] & $2.81\,[2.24\text{--}3.80]$ & $1.71\,[1.43\text{--}2.12]$ & $1.29\,[1.03\text{--}1.65]$ \\
$T$ [K] & $9432\,[8257\text{--}9910]$ & $10014\,[9389\text{--}10380]$ & $10250\,[9700\text{--}10619]$ \\
$N_{\rm H}$ [$10^{20}$ cm$^{-2}$] & $1.09\,[0.80\text{--}1.66]$ & $0.66\,[0.51\text{--}0.96]$ & $0.51\,[0.38\text{--}0.73]$ \\
\hline
\multicolumn{4}{l}{\emph{Mass-weighted}} \\
$R_{\rm eff}$ [pc] & $21.6\,[8.6\text{--}43.4]$ & $42.3\,[19.8\text{--}58.5]$ & $49.5\,[24.6\text{--}74.0]$ \\
$M$ [$10^{4}\,\msun$] & $1.99\,[0.05\text{--}8.41]$ & $8.14\,[0.41\text{--}16.55]$ & $9.36\,[1.26\text{--}18.42]$ \\
$n_{\rm H}$ [cm$^{-3}$] & $4.10\,[3.96\text{--}5.08]$ & $3.00\,[3.00\text{--}3.00]$ & $1.95\,[1.95\text{--}2.13]$ \\
$T$ [K] & $6290\,[5549\text{--}6497]$ & $5144\,[5096\text{--}5192]$ & $5213\,[5180\text{--}5246]$ \\
$N_{\rm H}$ [$10^{20}$ cm$^{-2}$] & $8.64\,[2.76\text{--}14.03]$ & $9.84\,[4.73\text{--}13.13]$ & $7.75\,[5.12\text{--}9.75]$ \\
\hline
\multicolumn{4}{l}{\emph{Survival criterion}} \\
$\chi$ & $444$ & $374$ & $344$ \\
$t_{\rm cool,mix}$ [Myr] & $0.042$ & $0.062$ & $0.077$ \\
$t_{\rm cc}$ [Myr] & $0.69$ & $0.71$ & $0.73$ \\
$r_{\rm crit}$ [pc] & $0.37$ & $0.54$ & $0.66$ \\
$R_{\rm eff}/r_{\rm crit}$ & $16.4$ & $11.5$ & $9.5$ \\
$\ell_{\rm cool}$ [pc] & $3.4$ & $5.6$ & $6.4$ \\
\hline
\end{tabular}
\end{table*}
To quantify the structure of the neutral clouds, we select five snapshots that are roughly uniformly-spaced in time from $49-70.6$ Myr, tracking the period of maximum neutral outflow rate. In each snapshot, we select for analysis three slabs of thickness 600 pc, chosen to provide sufficient statistical sampling, covering heights $1.2-1.8$ kpc, $2.2 - 2.8$ kpc, and $3.2-3.8$ kpc. Within each slab, we define a neutral cloud as a connected set of cells satisfying $x_e < 0.05$, where connection is evaluated keeping in mind that the domain is periodic in the horizontal direction. We discard objects containing fewer than 500 cells (an effective radius of $4.7$ pc) as unresolved, and use the remainder for our analysis.

For each identified cloud, we compute the following statistics: total mass $M$, effective radius $R_\mathrm{eff}$ (defined as the radius of a sphere whose volume is equal to the cloud volume), hydrogen number density $n_\mathrm{H}$ (defined as a volume-weighted mean), temperature $T$ (a mass-weighted mean), and equivalent column density $N_\mathrm{H}$ (defined as the line of sight hydrogen column density through the centre of a spherical cloud of equal mass and volume). Quantitatively, for a set of $N_c$ cells belonging to a given cloud, we define
\begin{align}
    M & = \Delta x^3 \sum_{i=1}^{N_c} \rho_i \\
    R_\mathrm{eff} & = \left(\frac{3}{4\pi} N_c \Delta x^3\right)^{1/3} \\
    n_\mathrm{H} & = \frac{1}{N_c\,\mu_\mathrm{H}} \sum_{i=1}^{N_c} \rho_i \\
    T & = \frac{1}{M} \Delta x^3 \sum_{i=1}^{N_c} \rho_i T_i \\
    N_\mathrm{H} & = 2 R_\mathrm{eff} \left(\frac{3M}{4\pi R_\mathrm{eff}^3 \mu_{\rm H}}\right),
\end{align}
where $\rho_i$ and $T_i$ are the total density and temperature in cell $i$ and $\mu_{\rm H} = 2.3 \times 10^{-24}$ g is the mean mass per H nucleon for our adopted chemical composition.

Finally, we compute the following statistics for the cloud population as a whole in each slab: the number density of clouds $n_\mathrm{cl}$ (averaged over the five time samples), fraction of cloud mass $f_M(R_\mathrm{eff} \geq 30\,\mathrm{pc})$ held in clouds with $R_\mathrm{eff} \geq 30$~pc (again averaged over the five samples), and the median and inter-quartile range of each of the per-cloud statistics, computed both by number and by mass.

We report these quantities for the 1 pc resolution run in Table~\ref{tab:clouds}. The Table reveals that the neutral material is arranged into a population of discrete structures with a broad spectrum of sizes and masses: most of the neutral mass is in structures with effective radii $R_\mathrm{eff} \sim 10$--$70$~pc, most structures by number have $R_\mathrm{eff} \approx 5$--$8$~pc, close to our detection limit of $4.7$~pc, and masses span $M \sim 10^2$--$10^5\,\msun$; typical column densities are $N_\mathrm{H}\sim 10^{20}-10^{21}$ cm$^{-2}$. These figures are in broad agreement with the properties observed for clouds in the Milky Way's nuclear wind \citep{DiTeodoro2018, Lockman2020, Noon2023}. Our neutral gas is, however, slower than those clouds: averaged over $29$--$112$~Myr, its mass-weighted mean velocity rises from $\approx 50\,\kmps$ at $1$~kpc to $\approx 130\,\kmps$ at $4$~kpc (Fig.~\ref{fig:ladder}), whereas \citet{Lockman2020} infer outflow velocities that rise from $\approx 150$--$200$ to $\approx 330\,\kmps$ at $2.5$--$3.5$~kpc. This difference may arise because the observed clouds are in the Galactic-centre wind, which is a different environment from our disc patch. Cloud densities and temperatures are typical of warm neutral ISM, $n_\mathrm{H} \sim \mathrm{few}$ cm$^{-3}$ and $T \sim 5000-10,000$ K. The mass distribution includes many small clouds, which tend to be somewhat less dense and hotter, but much of the mass is in large clouds, and the dominance of larger clouds grows as the clouds move away from the plane. We quantify this trend by measuring the fraction of the mass in structures with effective radii $> 30$ pc, which grows from $\approx 50\%$ at $|z|\approx 1.5$ kpc to $\approx 70\%$ at $|z| \approx 3.5$ kpc; the largest clouds have filamentary morphologies, with the cloud shown in the inset of Fig.~\ref{fig:morph} being a typical example.

\subsection{Effects of resolution}
\label{sec:resolution}

\begin{figure*}
\centering
\includegraphics[width=\textwidth,height=0.8\textheight,keepaspectratio]{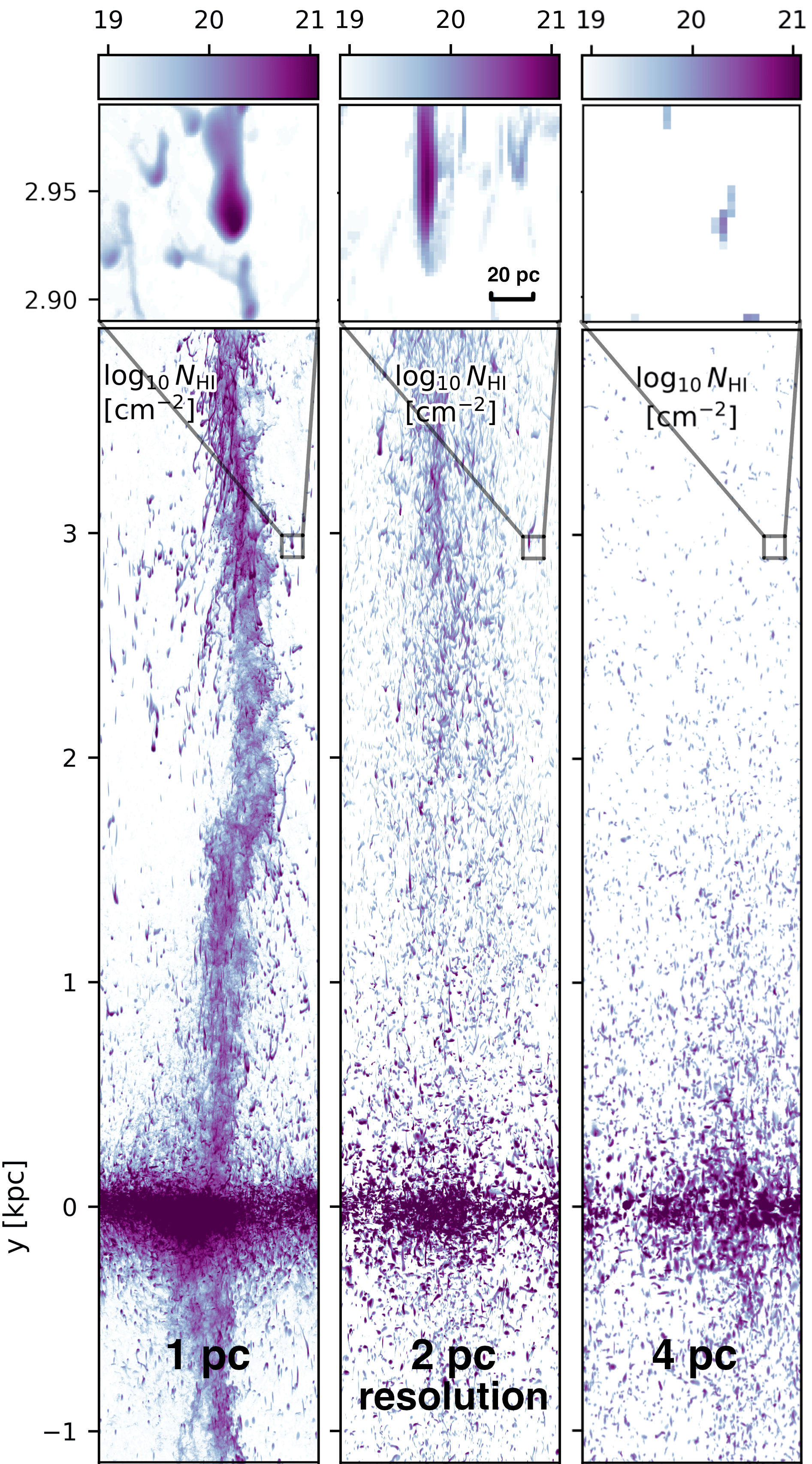}
\caption{Neutral gas column density projections at different simulation resolutions. The three panels show the projected neutral column at $t=88$ Myr at our three resolutions of $\Delta x= 1$ pc (left column), 2 pc (middle column), and 4 pc (right column). The insets show the same region as the insets in Fig.~\ref{fig:morph}. For clarity we show only a portion of the full simulation domain in these figures. The colour scale saturates at $N_{\him} = 10^{21}$~cm$^{-2}$, below the columns through the disc, so near the mid-plane it shows the area covered by dense neutral gas rather than its mass. At $4$~pc the neutral gas in the disc is concentrated in a thin layer of dense clumps, so it covers less of the map, although the $4$~pc box holds more neutral gas than the other two (Section~\ref{sec:resolution}).}
\label{fig:res}
\end{figure*}

Coarsening the grid weakens the neutral outflow at $2$~pc and removes it at $4$~pc. Figure~\ref{fig:res} shows the three runs at the same time. We see that at $4$~pc neutral gas in the wind is absent or barely resolved, while at $2$~pc it is present but is significantly less prominent, and has a qualitatively different spatial distribution than at our maximum resolution. The neutral mass loading factors confirm this visual impression: at the box edge we find $\eta_{\him}$ and $\eta_{\him}/\eta_\mathrm{total}$ are both $\lesssim 10^{-3}$ with $\Delta x = 4$ pc, $\eta_{\him} \approx 0.12$ and $\eta_{\him}/\eta_\mathrm{total} \approx 0.3$ with $\Delta x = 2$ pc, compared to $\eta_{\him} \approx 0.3$ and $\eta_{\him}/\eta_\mathrm{total} \approx 0.4$ at the full $1$~pc resolution; see Table~\ref{tab:ladder} for precise figures. Thus a factor of two reduction in resolution decreases the neutral mass loading by a factor of $\approx 2.7$ while a factor of four reduction makes it disappear almost entirely. Examining how $\eta_{\him}$ changes with height sharpens this picture: at $4$~pc resolution, there is no net neutral outflow at any height above $1$~kpc, whereas at $\Delta x = 2$ and $1$~pc the same quantity varies by less than $7$ and $19$ per cent from $1$ to $4$~kpc. The total mass loading also changes little with height, by $2$--$19$ per cent at every resolution, but from $\Delta x = 1$ to $4$~pc it falls by a factor of $12$, whereas $\eta_{\him}$ falls by more than two orders of magnitude. Coarsening the grid therefore suppresses the neutral outflow much more strongly than the total outflow.

\begin{figure}
\centering
\includegraphics[width=\columnwidth]{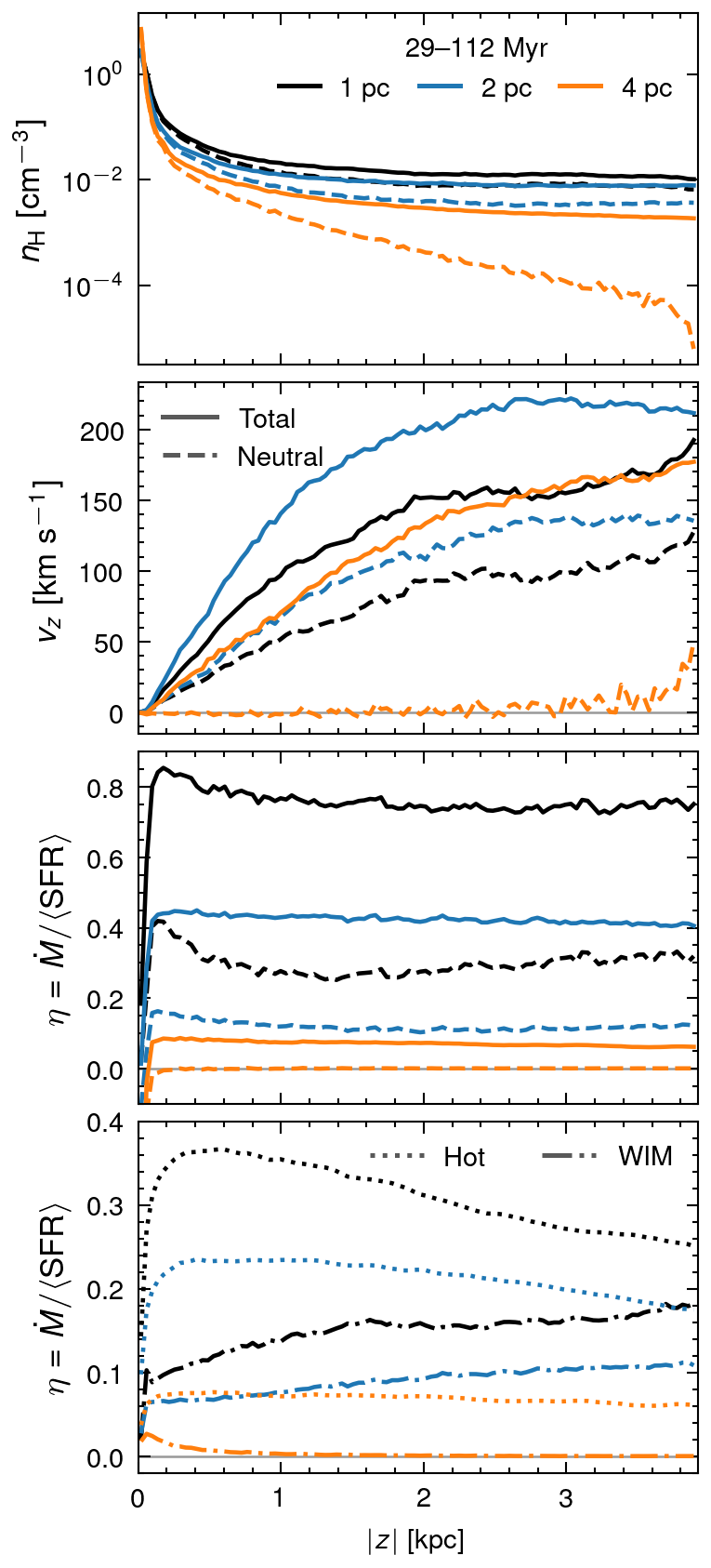}
\caption{Time-averaged vertical profiles at the three resolutions, $\Delta x = 1$~pc (black), $2$~pc (blue) and $4$~pc (orange), averaged over $29$--$112$~Myr. From top to bottom: hydrogen number density $n_{\rm H}$, mass-weighted mean outward velocity $v_z$, and mass loading factor $\eta$ of the total (solid) and neutral (dashed) gas, and $\eta$ of the hot gas (dotted) and the WIM (dash-dotted). Profiles are binned in $|z|$ with a bin width of $39$~pc. The $\eta$ axis of the third panel is clipped at $-0.1$; the $4$~pc total and neutral curves reach $\eta \approx -0.6$ in the first bin, inside the disc. The $2$ and $4$~pc runs do not record the density of the hot gas, so the hot gas and the WIM are shown by their mass flux only.}
\label{fig:ladder}
\end{figure}

Figure~\ref{fig:ladder} shows time-averaged vertical profiles of density, velocity and mass loading at all three resolutions. The absence of a neutral outflow at $4$~pc is not due to a lack of neutral gas. At every output between $29$ and $112$~Myr, the $4$~pc box contains $1.2$--$1.6$ times as much neutral gas as the $1$~pc box ($2.1\times10^{7}$ against $1.5\times10^{7}\,\msun$ on average, and $1.2\times10^{7}\,\msun$ at $2$~pc). Part of this difference comes from the star formation before the interval: by $29$~Myr the $4$~pc run had formed $0.9\times10^{7}\,\msun$ of stars, against $1.7\times10^{7}\,\msun$ at $1$~pc. However, this gas stays in the disc plane. On average only $4$ per cent of the neutral mass lies above $|z| = 0.1$~kpc and $0.4$ per cent above $1$~kpc, compared to $21$ and $6.5$ per cent at $2$~pc and $33$ and $11$ per cent at $1$~pc; and at no output does the fraction above $0.1$~kpc exceed $9$ per cent at $4$~pc, whereas it reaches $43$ and $55$ per cent at $2$ and $1$~pc. The mass outflow rate shows the same behaviour. At $|z| = 0.1$~kpc, just above the disc, the net neutral flux at $4$~pc is close to zero ($\eta_{\him} \approx -0.02$), compared to $\eta_{\him} \approx 0.16$ at $2$~pc and $0.40$ at $1$~pc, and the small amount of neutral gas above $1$~kpc has a mean vertical velocity close to zero. Thus at $4$~pc resolution a large amount of neutral gas exists in the simulation box, but it stays in the disc plane and is not launched above the disc. Instead, it is used up by star formation: over the interval the neutral mass in the $4$~pc box falls by $3.0\times10^{7}\,\msun$, while the run forms $3.0\times10^{7}\,\msun$ of stars; at $1$~pc the neutral mass falls by $2.4\times10^{7}\,\msun$ while the run forms $1.5\times10^{7}\,\msun$ of stars. Our runs cannot tell which comes first: the neutral gas may collapse into dense clumps and form stars because it is not launched, or it may be consumed by star formation before it can be launched.

The total outflow also collapses at $4$~pc, and for the same reason: little gas of any phase is launched. At $4$~pc, $\eta_\mathrm{total} = 0.06$--$0.08$ at every height from $0.1$ to $4$~kpc, compared to $\approx 0.4$ at $2$~pc and $0.75$--$0.85$ at $1$~pc, and only $2.5$ per cent of the gas mass in the box lies above $1$~kpc, compared to $12$ and $16$ per cent at $2$ and $1$~pc. Almost all of the $4$~pc outflow is hot gas, but the hot outflow is also weaker: at $|z| = 0.1$~kpc, $\eta_\mathrm{hot} = 0.07$ at $4$~pc, compared to $0.20$ at $2$~pc and $0.31$ at $1$~pc, although the $4$~pc run has the highest star formation rate. This weaker hot outflow indicates a difference in the supernova recipe rather than in the wind. At $\Delta x = 4$~pc the Sedov--Taylor phase of the remnants is more often unresolved, so the momentum-energy switch (Section~\ref{sec:numerics}) more often picks the momentum branch, which deposits the terminal momentum of the remnant rather than its energy and therefore produces less hot gas. We caution that this also reduces the amount of gas that the $4$~pc run lifts out of the disc, so the $4$~pc result reflects both the resolution of the wind and this change in the supernova recipe. The WIM behaves differently from the hot gas. At $2$ and $1$~pc, $\eta_\mathrm{WIM}$ increases with height, from $0.07$ and $0.09$ at $0.1$~kpc to $0.11$ and $0.17$ at the box edge, while the hot mass flux decreases above $0.5$~kpc. This is consistent with hot gas that cools into the WIM in the boundary layers of the neutral clouds (Section~\ref{sec:accel}). At $4$~pc, where there are no neutral clouds above the disc, $\eta_\mathrm{WIM}$ instead decreases from $0.02$ at $0.1$~kpc to zero at the box edge. Thus the collapse of $\eta_\mathrm{total}$ at $4$~pc is set at launch: the coarse run lifts much less gas of every phase out of the disc, and without neutral clouds aloft there is also no WIM produced around them.

\begin{figure}
\centering
\includegraphics[width=\columnwidth]{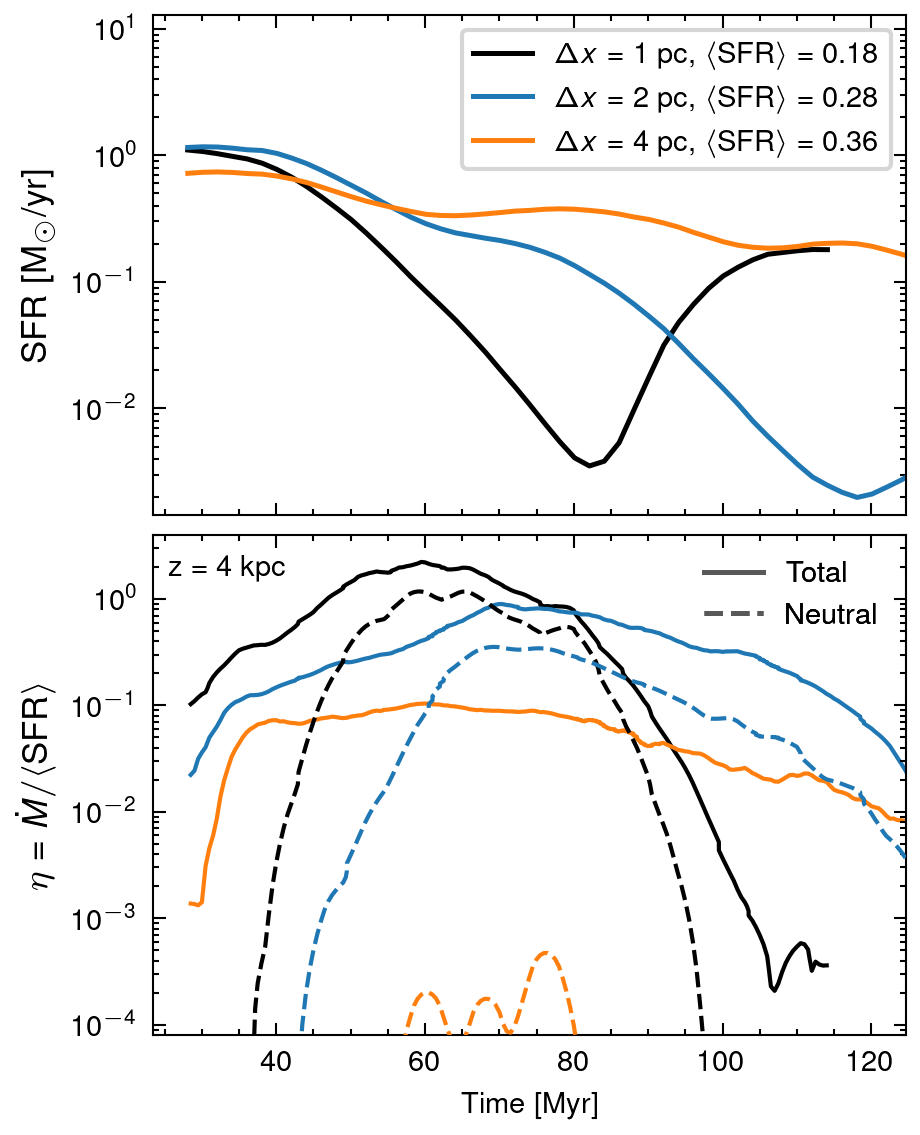}
\caption{Star formation rate (top) and mass loading factor (bottom) versus time in the three simulations. The mass-loading factor is measured at the top of the simulation domain, $z = 4$ kpc, and we show both the total mass-loading including all phases (solid) and the component carried by the neutral phase alone (dashed), both normalised by the mean star formation rate over $29$--$112$~Myr (Table~\ref{tab:runs}).}
\label{fig:time}
\end{figure}

The difference in the behaviour of the neutral outflow is not due to a difference in the amount of feedback. We illustrate this in Fig.~\ref{fig:time}, which shows the star formation rate (top panel) and both the total and neutral mass loading as a function of time. The figure shows that the mean star formation rate differs by a factor of $\approx 2$ between the resolutions, but the $1$~pc run, which has the most neutral outflow, has the lowest star formation rate (Table~\ref{tab:runs}). The total mass loading factor $\eta_\mathrm{total}$ differs by a factor of $\approx 1.8$ between 1 and 2 pc resolution, and by a factor of $\approx 7$ from 2 pc to 4 pc, compared to a difference of more than two orders of magnitude in the neutral mass loading. The feedback budget therefore does not explain the trend, although at $4$~pc the way in which the feedback couples to the gas does change, as discussed above.

\subsection{Acceleration of the neutral gas}
\label{sec:accel}

\begin{figure}
\centering
\includegraphics[width=\columnwidth]{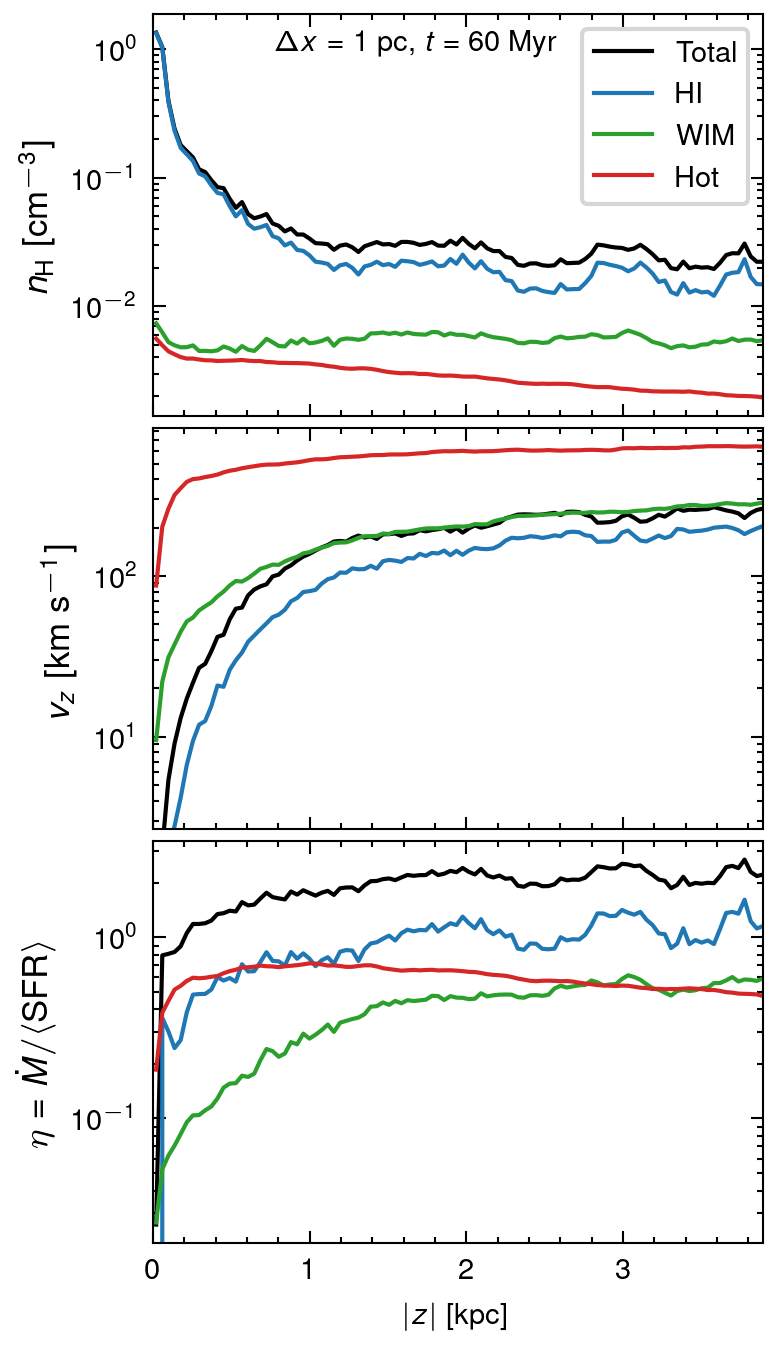}
\caption{Vertical structure of the outflow at $t=60$ Myr. The panels show, from top to bottom, the mean density of the gas expressed as H nucleons per unit volume $n_\mathrm{H}$, the mass-weighted mean vertical velocity $v_z$, and the mass loading factor $\eta$ as a function of height $|z|$; all quantities are averages over slabs of thickness $\Delta z \approx 39$~pc, combining the slabs at $+z$ and $-z$. Black shows the sum over all phases, while blue, green, and red show the neutral, WIM, and hot phases, respectively. The mass loading is normalised by the mean star formation rate over $29$--$112$~Myr (Table~\ref{tab:runs}).}
\label{fig:zprof}
\end{figure}

In order to better understand why the neutral gas survives and how it is accelerated into the wind in our high-resolution simulation, and why this does not occur at lower resolution, we must investigate its structure and momentum budget. In Fig.~\ref{fig:zprof} we show the vertical profile of gas density, mass-weighted mean vertical velocity, and mass loading for the $\Delta x = 1$ pc simulation at $t=60$ Myr. We see that neutral gas dominates the mass at every height above the disc, and the mass loading above $|z| \approx 1$~kpc, tracking the total to within a factor of two, with the hot phase about an order of magnitude less dense and the warm ionised medium between them. The hot gas velocity rises steeply near the disc, reaching $\approx 400$ km s$^{-1}$ within a few hundred pc of the midplane and $600-700$ km s$^{-1}$ by the time it reaches the edge of the simulation domain, while the neutral velocity rises more slowly, not passing $100\,\kmps$ until heights of $\approx 1$ kpc, and reaching $\sim200\,\kmps$ only at the top of the box.

\begin{figure}
\centering
\includegraphics[width=\columnwidth]{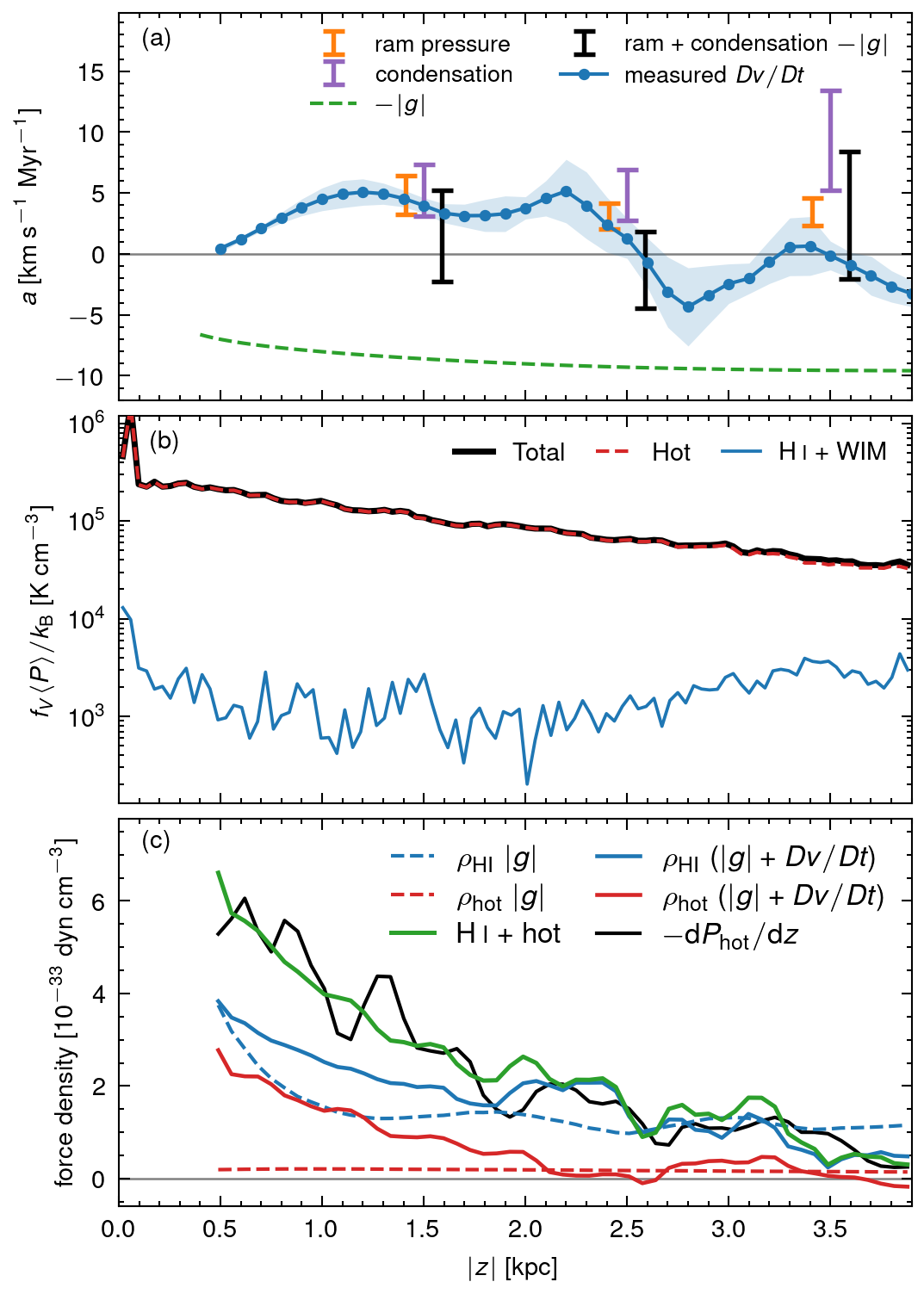}
\caption{Neutral gas acceleration and momentum budget. \textit{(a)} Measured Lagrangian acceleration of the neutral gas (blue, with the shaded band its uncertainty), the gravitational acceleration (green dashed), the momentum transferred from the hot phase to the neutral phase by ram pressure and condensation at $z = 1.5, 2.5$, and 3.5 kpc (orange and purple bars, respectively), and the sum of ram pressure, condensation, and gravity (black bars) at heights $z\approx 1.5$, $2.5$, and $3.5$ kpc. The ram pressure bar spans $C_D=1$ and $2$, and the condensation bar spans an arrival velocity from the boundary-layer contrast to the hot-phase mean. \textit{(b)} Volume-filling thermal pressure: total (black), hot (red dashed), and the rest of the gas, H~\textsc{i} plus WIM (blue). \textit{(c)} Force density per phase, H~\textsc{i} (blue), hot (red) and their sum (green); solid lines give the total force required to explain the measured acceleration, while dashed lines show the force required to counter gravity. The solid black line is $-\mathrm{d}P_\mathrm{hot}/\mathrm{d}z$, the force provided by the hot phase. Accelerations, pressures, and forces are all measured at $t = 60$ Myr, while the ram pressure and condensation points average over five snapshots spanning $49-70.6$ Myr; see Appendices~\ref{sec:lagrangian} and \ref{sec:drag}.
}
\label{fig:support}
\end{figure}

A velocity that rises with height does not by itself mean the gas is accelerating, since the same sort of pattern could be produced by a Hubble-type flow where the highest velocity material is at the largest distances simply because it was launched fastest. To establish that the neutral gas is in fact accelerating, we estimate the Lagrangian derivative of the neutral velocity as described in Appendix~\ref{sec:lagrangian}, and plot the result in Fig.~\ref{fig:support}a. We see that the neutral gas is in fact accelerating upwards at $\sim 5\,\accunit$ out to distances of $\approx 2.5$ kpc from the disc, after which point it coasts at near constant velocity. By contrast, gas travelling ballistically under the influence of gravity would be expected to \emph{decelerate} at $|g|\approx 7$--$10\,\accunit$ over the same range of heights (dashed green line in Fig.~\ref{fig:support}a); indeed, a parcel launched ballistically anywhere below $1.0$~kpc at the mean neutral gas speed we measure there would stall and turn around before reaching the top of the simulation domain.

The reservoir of pressure, and thus of momentum flux, that can counteract gravity is the hot phase, which carries $\approx 99\%$ of the volume-filling pressure $f_V\langle P\rangle$ (Fig.~\ref{fig:support}b); here we define $f_V\langle P\rangle$ as the product of a phase's volume filling fraction and its mean pressure, so that the phases sum exactly to the total. The hot phase pressure falls monotonically from $f_V\langle P\rangle/k_B\approx 2\times10^{5}$ to $5.0\times10^{4}\,\mathrm{K\,cm^{-3}}$ between $0.5$ and $3$~kpc, sustaining an outward gradient across the whole box.

This gradient is much larger than would be needed to support the hot gas against gravity, or even to accelerate it, implying that the decrease in pressure of the hot phase is caused by a transfer of momentum to other phases. To demonstrate this, note that the rate of change of momentum per unit volume, or equivalently the force density, applied to the gas in phase $p$ is $\rho_p(Dv_p/Dt + g)$; when summed over all three phases, this must equal the decrease in the available momentum flux in the hot gas pressure, $-\mathrm{d}P_{\rm hot}/\mathrm{d}z$. We evaluate these various terms as described in Appendix~\ref{sec:lagrangian}, and show the result in Fig.~\ref{fig:support}c. The figure shows that, integrated over $0.5$--$3$~kpc, the neutral phase takes up $2.5$ times as much momentum as the hot phase does, and the two together fully account for the decrease in hot gas pressure to within $\sim 30\%$ -- very good agreement given the approximations that are necessary to compute the various terms. The neutral phase dominates the uptake of momentum because, while it moves at a third of the hot velocity, it carries seven times the mass.

The remaining question then becomes how the hot gas transfers momentum to the neutral phase, for which there are two plausible channels. One is ram pressure combined with drag: as the hot gas flows past the cold phase, hot material impacting the surface of cold clouds and the boundary layers around them exerts a force that transmits momentum. The other is condensation: if hot gas cools into the neutral phase, it will carry its higher momentum with it, thereby increasing the mean momentum of the cool gas; this channel entrains cold clouds in wind-tunnel simulations \citep{GronkeOh2018, GronkeOh2020}, and a similar transfer of momentum by mixing accelerates the cool gas in a global simulation of a starburst-driven wind \citep{Schneider2020}. Including both sources of acceleration, and the effects of gravity, we can express the total acceleration experienced by a neutral cloud of mass $M$ and cross-sectional area $A$ as
\begin{equation}
\frac{Dv}{Dt} = -|g| + \frac{1}{M}\left(\frac{1}{2} C_D A \rho_\mathrm{h} \Delta v^2
  + \Gamma_\mathrm{h\to\him}\Delta v\right),
\label{eq:budget}
\end{equation}
where $g$ is the gravitational acceleration, $C_D$ is a coefficient of drag of order unity, $\rho_\mathrm{h}$ is the density of the hot gas impacting the cloud, $\Delta v$ is the velocity difference between the hot wind and the neutral cloud, and $\Gamma_\mathrm{h\to\him}$ is the net rate at which hot gas condenses onto the cloud; the first term in parentheses represents ram pressure and drag, and the second represents condensation. The main challenge in evaluating these two terms is the choice of $\rho_\mathrm{h}$ and $\Delta v$, since these vary between the mean properties of the hot gas and the properties in a boundary layer around each neutral cloud. We consider both possibilities, and thereby estimate the mean value of both terms on the right hand side over all neutral clouds, as described in Appendix~\ref{sec:drag}. We show the range of estimates at which we arrive as the orange and purple bars in Fig.~\ref{fig:support}a.

The figure shows that ram pressure/drag and condensation contribute comparably at $1.5$~kpc, with condensation increasingly dominant higher up, and that neither suffices alone: even at $C_D = 2$ ram pressure reaches only $6.5\,\accunit$ against $|g| = 8.6\,\accunit$ at $1.5$~kpc, so drag can never lift the gas by itself at any height. Together, condensation and ram pressure are comparable to or stronger than gravity. The black bars in Fig.~\ref{fig:support}a show their sum minus gravity; at each height their range includes the measured acceleration, so the forces we have computed are consistent with the acceleration of the H~\textsc{i} clouds in our high-resolution simulation, although the uncertainty is as large as the signal.

\subsection{On the survival of neutral gas, and the importance of resolution}
\label{sec:survival}

Condensation also explains why H~\textsc{i} structures exist at all. To see this, we begin by estimating the crushing times for our neutral clouds, $t_{\rm cc} = \chi^{1/2}R_{\rm eff}/\Delta v$ \citep{Klein1994}, where $R_{\rm eff}$ is the effective cloud radius, $\Delta v$ is the velocity difference between the cloud and the hot material that is accelerating it, and $\chi = \rho_{\rm cl}/\rho_{\rm h}$ is the ratio of the cloud density to the density of the hot material pushing on it. We estimate each of the quantities entering $t_{\rm cc}$ in each of our sample slabs introduced in Section~\ref{sec:fullres}. To do so we first define a boundary layer around each cloud following the algorithm outlined in Appendix~\ref{sec:drag}. To compute $\chi$, we divide the number-weighted median densities of H~\textsc{i} clouds found in each slab by the number-weighted median densities of their boundary layers. We use the number-weighted medians rather than the mass-weighted medians because the former emphasise values typical of the small clouds that are most vulnerable to disruption by the wind. We report the resulting values of $\chi$ in Table~\ref{tab:clouds}; typical values are $\chi\approx 400$. We use a completely analogous method to estimate the velocity differences between clouds and boundary layers, finding a typical $\Delta v \approx 150$--$180\,\kmps$. Combining these figures with our typical $R_{\rm eff}$ values (also listed in Table~\ref{tab:clouds}) gives a typical $t_\mathrm{cc} \approx 0.7$~Myr at every height. This is significantly smaller than the flow time $z/v_z$ $\sim 15$~Myr. Thus neutral clouds are found at heights that a cloud launched from the disc would take roughly $20$ cloud-crushing times to reach. They may have been launched from the disc and survived this long, or they may have formed in the wind on the way; either way, neutral gas persists in the wind for many cloud-crushing times.

An adiabatic cloud would certainly have been destroyed by this time, but condensation can compensate for the mass loss caused by stripping and allow clouds to survive; \citet{GronkeOh2018} find that cold clouds survive if their boundary layers cool fast enough, with a quantitative requirement that the cooling time in the mixing layer satisfy $t_\mathrm{cool,mix} < t_\mathrm{cc}$, where the cooling time $t_\mathrm{cool}$ is defined as the ratio of the gas internal energy to the radiative cooling rate, and $t_\mathrm{cool,mix}$ is evaluated at temperature $T_\mathrm{mix} = \sqrt{T_\mathrm{h} T_\mathrm{cl}}$ and density $\rho_\mathrm{mix} = \sqrt{\rho_\mathrm{h} \rho_\mathrm{cl}}$, i.e., at the geometric mean temperature and density between conditions in the cloud and in the surrounding hot gas.

We evaluate $T_\mathrm{h}$, $T_\mathrm{cl}$, $\rho_\mathrm{h}$, and $\rho_\mathrm{cl}$ using the same number-weighted medians in the clouds and their boundary layers as used when evaluating $\chi$. Once these values are determined, we compute cooling times at $\rho_\mathrm{mix} = \sqrt{\rho_\mathrm{h} \rho_\mathrm{cl}}$ and $T_\mathrm{mix} = \sqrt{T_\mathrm{h} T_\mathrm{cl}}$ by reading the cooling rate at that combination of density and temperature from our \textsc{Grackle} cooling tables, and dividing this into the internal energy at the corresponding density and temperature. We report the values of $t_\mathrm{cool,mix}$ that we obtain by this procedure for each of our sample slab regions in Table~\ref{tab:clouds}.

In each of our three sample slabs we find cooling times $t_\mathrm{cool,mix} < 0.08$ Myr, roughly an order of magnitude smaller than the cloud crushing time.
Equivalently, the critical radius for cloud survival $r_{\rm crit} = \Delta v\,t_{\rm cool,mix}/\chi^{1/2}$ is $\sim 0.5$~pc, an order of magnitude below the median cloud effective radius $R_\mathrm{eff}$ (Table~\ref{tab:clouds}). This explains why we observe that our H~\textsc{i} clouds grow rather than being destroyed as the wind moves upward, as do the cool clouds in the tall-box simulations of \citet{TanFielding2024}, whose mass growth rates follow turbulent radiative mixing-layer theory.
We compare with other survival criteria in Section~\ref{sec:disc_res}.

This analysis also suggests why our $4$~pc simulation, like earlier simulations with similar resolution, has almost no H~\textsc{i} in the wind, and why our $2$~pc simulation has less than our $1$~pc one. The characteristic size of the cooling layer is $\ell_{\rm cool} = c_{\rm s}t_{\rm cool,mix}$, where $c_{\rm s}$ is the sound speed in the mixing layer (i.e., the sound speed evaluated at a temperature $T_\mathrm{mix}$; \citealt{GronkeOh2018}). Evaluating this for our values of $T_\mathrm{mix}$ for our sample slabs gives $\ell_{\rm cool} \approx 3-6$ pc at heights of 1.5 to 3.5 kpc (Table~\ref{tab:clouds}). In our $\Delta x= 1$ pc resolution simulation, these layers are at least marginally resolved by $\approx 3-6$ cells, but at a resolution $\Delta x = 2$ pc this becomes even more marginal, $\approx 1.5 - 3$ cells, and at $\Delta x = 4$ pc the mixing layer is reduced to a single cell, and even the typical neutral cloud ($R_\mathrm{eff} \sim 6$ pc) is only about three cells across. Consequently, our lower-resolution simulations fail to capture the mixing layers that are responsible for both half the momentum transfer from the hot phase to the H~\textsc{i}, and for all of the condensation that re-supplies the H~\textsc{i}. At $\Delta x = 4$~pc the change in the supernova recipe (Section~\ref{sec:resolution}) adds to this effect, because it reduces the amount of gas launched from the disc.

\section{Discussion}
\label{sec:discussion}

\subsection{{The resolution requirement for neutral winds}}
\label{sec:disc_res}

One strong implication of our work is that the steep vertical decline of the neutral phase reported by previous stratified-box and Galactic-centre calculations of supernova-driven outflows is, at least in part, numerical. Our own $4$~pc run reproduces that decline; the same physical model at $1$~pc does not. Our analysis of the boundary layers around the H~\textsc{i} clouds produced in our 1~pc resolution simulation provides a quantitative explanation for this phenomenon: in order to produce H~\textsc{i} accelerated into winds, simulations need to resolve the boundary layers responsible for both re-supplying their mass and transferring hot gas momentum to them. The characteristic sizes of these structures are a few pc, so a resolution of at least $\Delta x = 1$~pc is required: $\Delta x = 1$~pc resolves them with $3$--$6$ cells, although $\eta_{\him}$ is not yet converged there (Section~\ref{sec:limitations}), while $\Delta x = 4$~pc clearly does not; $\Delta x = 2$ pc is marginal, and the outcome likely depends on details of the adopted cooling curve and on the level of numerical dissipation in the simulation method. At $\Delta x = 4$~pc part of the decline also comes from the supernova recipe, which switches to the momentum branch more often and so launches less gas (Section~\ref{sec:resolution}).

The \citet{GronkeOh2018} criterion that we apply in Section~\ref{sec:survival} was calibrated with clouds at $T_{\rm cl} \approx 4\times10^4$~K, and it performs less well for colder clouds \citep{FarberGronke2022, Abruzzo2023}. For clouds near $10^4$~K, such as ours, \citet{FarberGronke2022} replace $t_{\rm cool,mix}$ by $t_{\rm cool,minmix}$, the cooling time at $T_{\rm minmix} = (T_{\rm min,cool}T_{\rm h})^{1/2}$, where $T_{\rm min,cool}$ is the temperature at which the isobaric cooling time is shortest between $T_{\rm cl}$ and $T_{\rm h}$, and require $t_{\rm cc} > t_{\rm cool,minmix}$. \citet{Abruzzo2023} compare $t_{\rm cool,minmix}$ with the shear time $t_{\rm shear} = R_{\rm eff}/\Delta v$ instead of $t_{\rm cc}$, and predict survival for $\alpha\,t_{\rm shear} > t_{\rm cool,minmix}$ with $\alpha \approx 7$, i.e.\ for $R_{\rm eff} > R_{\rm crit} = \Delta v\,t_{\rm cool,minmix}/\alpha$. With the same median cloud properties and cooling table as in Section~\ref{sec:survival}, we find $T_{\rm min,cool} \approx 1.6\times10^4$~K, $T_{\rm minmix} \approx 2.4$--$2.6\times10^5$~K and $t_{\rm cool,minmix} \approx 0.08$--$0.13$~Myr at heights of $1.5$--$3.5$~kpc. The median cloud therefore passes the \citet{FarberGronke2022} criterion by a factor of $6$--$8$, with a critical radius $R_{\rm crit,FG} = \Delta v\,t_{\rm cool,minmix}/\chi^{1/2} \approx 0.7$--$1.1$~pc, but the \citet{Abruzzo2023} criterion only by a factor of $2$--$3$, with $R_{\rm crit} \approx 2$--$3$~pc. This margin is comparable to the uncertainty in $\alpha$, so the small clouds that dominate the population by number are marginal, while the clouds that hold most of the neutral mass (the mass-weighted medians in Table~\ref{tab:clouds}) pass by a factor of $6$--$7$. For the median cloud, $R_{\rm crit}$ spans $2$--$3$ cells at $\Delta x = 1$~pc but less than one cell at $\Delta x = 4$~pc (and $R_{\rm crit,FG}$ about one cell at $\Delta x = 1$~pc), so this criterion also implies that a resolution of at least $1$~pc is needed to follow the survival of the clouds. One caveat is that \citet{Abruzzo2023} calibrated their criterion for supersonic winds, whereas our clouds move through the hot gas at $\Delta v \approx 150$--$250\,\kmps$, close to or below the sound speed of hot gas in pressure balance with the clouds, which is $\chi^{1/2}$ times the cloud sound speed, or $\approx 200$--$230\,\kmps$. 

\subsection{{Limitations}}
\label{sec:limitations}

Our calculations are unmagnetised and include no explicit thermal conduction, so of the three mechanisms that have been proposed to explain the survival of cool, neutral material in winds -- radiative cooling, conduction, and magnetic fields -- they contain only the first. This demonstrates that radiative cooling of the mixing layers is sufficient on its own, but the sign of the effect of the other two on $\eta_{\him}$ is not obvious, because each acts in competing ways. In wind-tunnel simulations, magnetic fields draped around a cloud suppress mixing, lengthen its life and raise the drag on it \citep{McCourt2015, Cottle2020}; by suppressing mixing, however, they may also reduce the condensation that grows our clouds and supplies about half of their momentum (Section~\ref{sec:accel}). Conduction evaporates cool gas, but it can also compress clouds into filaments that resist mixing \citep{Bruggen2016}. In the disc, magnetic fields supply part of the vertical support \citep{KimOstriker2018}, which our simulations lack, and so may change the star formation rate and the structure of the gas from which the wind is launched. We therefore do not know whether these processes would raise or lower $\eta_{\him}$, and our results are not a bound in either direction with respect to them. It is also unclear how including these processes would modify the resolution requirement we have found, since both magnetic fields and conduction can alter the thickness of the mixing layer.

In addition, although we have demonstrated a qualitative change in behaviour with resolution -- from $\eta_{\him} \sim 0$ to $\eta_{\him} \sim 0.3$ as we go from $\Delta x = 4$ pc to $\Delta x = 1$ pc, we have not achieved numerical convergence: $\eta_{\him}$ is noticeably larger for $\Delta x = 1$ pc than for $\Delta x = 2$ pc, and with only three resolutions we cannot predict its value at $\Delta x = 1/2$ pc. The trend with resolution need not be monotonic, because resolution changes not only the mixing layers but also how much gas the supernovae launch from the disc (Section~\ref{sec:resolution}). Our neutral mass loading at $\Delta x = 1$ pc is therefore the value at the highest resolution we can reach, not a converged value or a bound. Moreover, while we are able to analyse the statistical properties of the H~\textsc{i} clouds that form in our simulations, we are limited in our ability to follow the detailed evolution of individual objects by the cadence of our outputs, which is a factor of a few longer than the cloud-crushing time and more than an order of magnitude longer than the cooling time.

\subsection{{Implications for galaxy-scale wind models}}
\label{sec:implications}

Despite these limitations, our simulations' demonstration that supernova-driven winds can be dominated by neutral material even at distances many kpc above the galactic plane has important consequences that reach beyond stratified box simulations. Subgrid wind prescriptions in cosmological simulations are calibrated on resolved calculations of exactly this kind \citep[e.g.,][]{Smith24a}, so if those calculations lose their neutral phase numerically due to inadequate resolution, the prescriptions derived from them inherit the loss. The error is not a small correction: at $4$~pc resolution there is no net neutral outflow at any height above $1$~kpc, while at $1$~pc resolution the neutral phase carries about $40$ per cent of the mass flux at every height. A second consequence is that, because the neutral phase is fed by condensation as it travels, the phase composition of a wind at the halo boundary is not the composition at launch. Winds cannot be treated as a fixed mixture advected outward. Instead, the neutral fraction delivered to the circumgalactic medium is set en route rather than at the disc.

Our finding that neutral gas can be carried to several kiloparsecs by supernova feedback alone also means that some of the motivation for simulations emphasising cosmic rays or radiation pressure as wind drivers is reduced. Of course this does not mean that these processes do not occur, simply that it may not be necessary to tune uncertain parameters in models of these processes (e.g., cosmic ray diffusion coefficients) in an attempt to reproduce observations of neutral material entrained high into winds. Our findings may therefore lead to a revision of models for cosmic ray- and radiation pressure-driven winds as well.

\section{Conclusions}
\label{sec:conclusions}

{We have simulated a supernova-driven outflow from a stratified patch of a galactic disc with $\siggas = 50\,\msun\,\mathrm{pc}^{-2}$ in a $1\,\mathrm{kpc} \times 1\,\mathrm{kpc} \times \pm 4\,\mathrm{kpc}$ box, using the GPU-accelerated code \quokka\ at grid spacings $\Delta x = 1$, $2$ and $4$~pc. The simulations include self-gravity, self-consistent star formation and supernova feedback, and radiative cooling from \textsc{Grackle}. Our main conclusions are as follows.
\begin{enumerate}
\item At $\Delta x = 1$~pc neutral gas dominates the mass of the wind at every height we measure: above $|z| = 1$~kpc it holds $64$--$69$ per cent of the gas mass and carries $35$--$43$ per cent of the mass flux, and $\eta_{\him} \approx 0.3$ and $\eta_{\him}/\eta_\mathrm{total} \approx 0.4$ at all heights from $1$ to $4$~kpc. Weighted by its atomic fraction, atomic gas carries $41$--$54$ per cent of the mass flux at these heights ($\eta_{\rm atom} \approx 0.3$--$0.4$).
\item The neutral gas forms discrete clouds with effective radii $R_\mathrm{eff} \sim 10$--$70$~pc, masses $M \sim 10^2$--$10^5\,\msun$ and column densities $N_\mathrm{H} \sim 10^{20}$--$10^{21}$~cm$^{-2}$, in broad agreement with the clouds observed in the Milky Way's nuclear wind. The fraction of the neutral mass in clouds with $R_\mathrm{eff} > 30$~pc grows from $\approx 50\%$ at $|z| \approx 1.5$~kpc to $\approx 70\%$ at $|z| \approx 3.5$~kpc.
\item The neutral gas is not ballistic. It accelerates upwards at $\sim 5\,\accunit$ out to $\approx 2.5$~kpc and then coasts, whereas gravity alone would decelerate it at $7$--$10\,\accunit$. The momentum comes from the pressure gradient of the hot phase, and reaches the neutral gas through ram pressure, drag, and condensation of hot gas onto the neutral phase. No single channel suffices, but together they can balance or exceed gravity, and the net acceleration they imply is consistent with the measured one.
\item Neutral clouds exist at heights that a cloud launched from the disc would take roughly $20$ cloud-crushing times to reach. Whether they were launched from the disc and survived or formed in the wind, this is possible because their mixing layers cool quickly: $t_\mathrm{cool,mix} < 0.08$~Myr, roughly an order of magnitude below the cloud-crushing time, so condensation can replace the mass lost to stripping.
\item The neutral outflow depends strongly on resolution. At $\Delta x = 4$~pc there is no net neutral outflow at any height above $1$~kpc. The box contains more neutral gas than at $\Delta x = 1$~pc, but this gas stays in the disc plane and is not launched: only $0.4$ per cent of it lies above $|z| = 1$~kpc, compared to $11$ per cent at $\Delta x = 1$~pc. The total outflow is also ten times weaker, because the coarse run lifts less gas of every phase out of the disc; at this resolution the supernova recipe more often switches to the momentum branch, which produces less hot gas. At $\Delta x = 2$~pc, $\eta_{\him} \approx 0.12$. One likely contributing cause is the cooling length of the mixing layers, $\ell_\mathrm{cool} \approx 3$--$6$~pc, which is resolved by $\approx 3$--$6$ cells at $\Delta x = 1$~pc but by only a single cell at $\Delta x = 4$~pc. Simulations should resolve this length, which requires a resolution of at least $1$~pc, to capture the neutral phase of supernova-driven winds. We have not achieved convergence at $\Delta x = 1$~pc, so $\eta_{\him}$ at higher resolution remains uncertain.
\end{enumerate}}

\section*{Acknowledgements}

CCH, MRK, and AV acknowledge support from the Australian Research Council through Laureate Fellowship FL220100020. This work was supported by computing resources provided by the Pawsey Supercomputing Research Centre’s Setonix Supercomputer, with funding from the Australian Government, through award jh2, and by the INCITE program at the Oak Ridge Leadership Computing Facility of USA.

\section*{Data Availability}

\quokka\ is publicly available at \url{https://github.com/quokka-astro/quokka}. Raw simulation outputs are not publicly available due to their large size, but will be provided on request to the corresponding author.

\bibliographystyle{mnras}
\bibliography{references,ref1,ref2}

\begin{thebibliography}{}
\makeatletter
\relax
\def\mn@urlcharsother{\let\do\@makeother \do\$\do\&\do\#\do\^\do\_\do\%\do\~}
\def\mn@doi{\begingroup\mn@urlcharsother \@ifnextchar [ {\mn@doi@}
  {\mn@doi@[]}}
\def\mn@doi@[#1]#2{\def\@tempa{#1}\ifx\@tempa\@empty \href
  {http://dx.doi.org/#2} {doi:#2}\else \href {http://dx.doi.org/#2} {#1}\fi
  \endgroup}
\def\mn@eprint#1#2{\mn@eprint@#1:#2::\@nil}
\def\mn@eprint@arXiv#1{\href {http://arxiv.org/abs/#1} {{\tt arXiv:#1}}}
\def\mn@eprint@dblp#1{\href {http://dblp.uni-trier.de/rec/bibtex/#1.xml}
  {dblp:#1}}
\def\mn@eprint@#1:#2:#3:#4\@nil{\def\@tempa {#1}\def\@tempb {#2}\def\@tempc
  {#3}\ifx \@tempc \@empty \let \@tempc \@tempb \let \@tempb \@tempa \fi \ifx
  \@tempb \@empty \def\@tempb {arXiv}\fi \@ifundefined
  {mn@eprint@\@tempb}{\@tempb:\@tempc}{\expandafter \expandafter \csname
  mn@eprint@\@tempb\endcsname \expandafter{\@tempc}}}

\bibitem[\protect\citeauthoryear{Abruzzo, Fielding  \& Bryan}{Abruzzo
  et~al.}{2023}]{Abruzzo2023}
Abruzzo M.~W.,  Fielding D.~B.,   Bryan G.~L.,  2023, preprint (\mn@eprint
  {arXiv} {2307.03228})

\bibitem[\protect\citeauthoryear{{Armillotta}, {Ostriker}, {Kim}  \&
  {Jiang}}{{Armillotta} et~al.}{2024}]{Armillotta24a}
{Armillotta} L.,  {Ostriker} E.~C.,  {Kim} C.-G.,   {Jiang} Y.-F.,  2024,
  \mn@doi [\apj] {10.3847/1538-4357/ad1e5c}, \href
  {https://ui.adsabs.harvard.edu/abs/2024ApJ...964...99A} {964, 99}

\bibitem[\protect\citeauthoryear{{Br{\"u}ggen} \& {Scannapieco}}{{Br{\"u}ggen}
  \& {Scannapieco}}{2016}]{Bruggen2016}
{Br{\"u}ggen} M.,  {Scannapieco} E.,  2016, \mn@doi [\apj]
  {10.3847/0004-637X/822/1/31}, \href
  {https://ui.adsabs.harvard.edu/abs/2016ApJ...822...31B} {822, 31}

\bibitem[\protect\citeauthoryear{{Chabrier}}{{Chabrier}}{2003}]{Chabrier2003}
{Chabrier} G.,  2003, \mn@doi [\pasp] {10.1086/376392}, \href
  {https://ui.adsabs.harvard.edu/abs/2003PASP..115..763C} {115, 763}

\bibitem[\protect\citeauthoryear{{Choi}, {Dotter}, {Conroy}, {Cantiello},
  {Paxton}  \& {Johnson}}{{Choi} et~al.}{2016}]{Choi16a}
{Choi} J.,  {Dotter} A.,  {Conroy} C.,  {Cantiello} M.,  {Paxton} B.,
  {Johnson} B.~D.,  2016, \mn@doi [\apj] {10.3847/0004-637X/823/2/102}, \href
  {http://adsabs.harvard.edu/abs/2016ApJ...823..102C} {823, 102}

\bibitem[\protect\citeauthoryear{Colella \& Woodward}{Colella \&
  Woodward}{1984}]{Colella1984}
Colella P.,  Woodward P.~R.,  1984, \mn@doi [J. Comput. Phys.]
  {10.1016/0021-9991(84)90143-8}, 54, 174

\bibitem[\protect\citeauthoryear{{Cottle}, {Scannapieco}, {Br{\"u}ggen},
  {Banda-Barrag{\'a}n}  \& {Federrath}}{{Cottle} et~al.}{2020}]{Cottle2020}
{Cottle} J.,  {Scannapieco} E.,  {Br{\"u}ggen} M.,  {Banda-Barrag{\'a}n} W.,
  {Federrath} C.,  2020, \mn@doi [\apj] {10.3847/1538-4357/ab76d1}, \href
  {https://ui.adsabs.harvard.edu/abs/2020ApJ...892...59C} {892, 59}

\bibitem[\protect\citeauthoryear{{Di Teodoro}, {McClure-Griffiths}, {Lockman},
  {Denbo}, {Endsley}, {Ford}  \& {Harrington}}{{Di Teodoro}
  et~al.}{2018}]{DiTeodoro2018}
{Di Teodoro} E.~M.,  {McClure-Griffiths} N.~M.,  {Lockman} F.~J.,  {Denbo}
  S.~R.,  {Endsley} R.,  {Ford} H.~A.,   {Harrington} K.,  2018, \mn@doi [\apj]
  {10.3847/1538-4357/aaad6a}, \href
  {https://ui.adsabs.harvard.edu/abs/2018ApJ...855...33D} {855, 33}

\bibitem[\protect\citeauthoryear{{Di Teodoro}, {McClure-Griffiths}, {Lockman}
  \& {Armillotta}}{{Di Teodoro} et~al.}{2020}]{DiTeodoro2020}
{Di Teodoro} E.~M.,  {McClure-Griffiths} N.~M.,  {Lockman} F.~J.,
  {Armillotta} L.,  2020, \mn@doi [\nat] {10.1038/s41586-020-2595-z}, \href
  {https://ui.adsabs.harvard.edu/abs/2020Natur.584..364D} {584, 364}

\bibitem[\protect\citeauthoryear{{Emerick}, {Bryan}  \& {Mac Low}}{{Emerick}
  et~al.}{2019}]{Emerick2019}
{Emerick} A.,  {Bryan} G.~L.,   {Mac Low} M.-M.,  2019, \mn@doi [\mnras]
  {10.1093/mnras/sty2689}, \href
  {https://ui.adsabs.harvard.edu/abs/2019MNRAS.482.1304E} {482, 1304}

\bibitem[\protect\citeauthoryear{{Farber} \& {Gronke}}{{Farber} \&
  {Gronke}}{2022}]{FarberGronke2022}
{Farber} R.~J.,  {Gronke} M.,  2022, \mn@doi [\mnras] {10.1093/mnras/stab3412},
  \href {https://ui.adsabs.harvard.edu/abs/2022MNRAS.510..551F} {510, 551}

\bibitem[\protect\citeauthoryear{{Fuchs}, {Jahrei{\ss}}  \& {Flynn}}{{Fuchs}
  et~al.}{2009}]{Fuchs09a}
{Fuchs} B.,  {Jahrei{\ss}} H.,   {Flynn} C.,  2009, \mn@doi [\aj]
  {10.1088/0004-6256/137/1/266}, \href
  {http://adsabs.harvard.edu/abs/2009AJ....137..266F} {137, 266}

\bibitem[\protect\citeauthoryear{{Girichidis} et~al.,}{{Girichidis}
  et~al.}{2016}]{Girichidis2016}
{Girichidis} P.,  et~al., 2016, \mn@doi [\mnras] {10.1093/mnras/stv2742}, \href
  {https://ui.adsabs.harvard.edu/abs/2016MNRAS.456.3432G} {456, 3432}

\bibitem[\protect\citeauthoryear{{Gronke} \& {Oh}}{{Gronke} \&
  {Oh}}{2018}]{GronkeOh2018}
{Gronke} M.,  {Oh} S.~P.,  2018, \mn@doi [\mnras] {10.1093/mnrasl/sly131},
  \href {https://ui.adsabs.harvard.edu/abs/2018MNRAS.480L.111G} {480, L111}

\bibitem[\protect\citeauthoryear{{Gronke} \& {Oh}}{{Gronke} \&
  {Oh}}{2020}]{GronkeOh2020}
{Gronke} M.,  {Oh} S.~P.,  2020, \mn@doi [\mnras] {10.1093/mnras/stz3332},
  \href {https://ui.adsabs.harvard.edu/abs/2020MNRAS.492.1970G} {492, 1970}

\bibitem[\protect\citeauthoryear{{Haardt} \& {Madau}}{{Haardt} \&
  {Madau}}{2012}]{Haardt2012}
{Haardt} F.,  {Madau} P.,  2012, \mn@doi [\apj] {10.1088/0004-637X/746/2/125},
  \href {https://ui.adsabs.harvard.edu/abs/2012ApJ...746..125H} {746, 125}

\bibitem[\protect\citeauthoryear{He, Wibking  \& Krumholz}{He
  et~al.}{2024a}]{He2024a}
He C.-C.,  Wibking B.~D.,   Krumholz M.~R.,  2024a, \mn@doi [\mnras]
  {10.1093/mnras/stae1244}, 531, 1228

\bibitem[\protect\citeauthoryear{He, Wibking  \& Krumholz}{He
  et~al.}{2024b}]{He2024b}
He C.-C.,  Wibking B.~D.,   Krumholz M.~R.,  2024b, \mn@doi [\mnras]
  {10.1093/mnras/stae2580}, 535, 3059

\bibitem[\protect\citeauthoryear{{He}, {Wibking}, {Vijayan}, {Krumholz}  \&
  {Li}}{{He} et~al.}{2026}]{He2026}
{He} C.-C.,  {Wibking} B.~D.,  {Vijayan} A.,  {Krumholz} M.~R.,   {Li} P.~S.,
  2026, \mn@doi [The Open Journal of Astrophysics] {10.33232/001c.159235},
  \href {https://ui.adsabs.harvard.edu/abs/2026OJAp....959235H} {9, 59235}

\bibitem[\protect\citeauthoryear{Kim \& Ostriker}{Kim \&
  Ostriker}{2017}]{Kim2017}
Kim C.-G.,  Ostriker E.~C.,  2017, \mn@doi [\apj] {10.3847/1538-4357/aa8599},
  846, 133

\bibitem[\protect\citeauthoryear{{Kim} \& {Ostriker}}{{Kim} \&
  {Ostriker}}{2018}]{KimOstriker2018}
{Kim} C.-G.,  {Ostriker} E.~C.,  2018, \mn@doi [\apj]
  {10.3847/1538-4357/aaa5ff}, \href
  {https://ui.adsabs.harvard.edu/abs/2018ApJ...853..173K} {853, 173}

\bibitem[\protect\citeauthoryear{{Kim} et~al.,}{{Kim}
  et~al.}{2020}]{KimSMAUG2020}
{Kim} C.-G.,  et~al., 2020, \mn@doi [\apj] {10.3847/1538-4357/aba962}, \href
  {https://ui.adsabs.harvard.edu/abs/2020ApJ...900...61K} {900, 61}

\bibitem[\protect\citeauthoryear{{Kim}, {Kim}, {Gong}  \& {Ostriker}}{{Kim}
  et~al.}{2023}]{Kim2023NCR}
{Kim} C.-G.,  {Kim} J.-G.,  {Gong} M.,   {Ostriker} E.~C.,  2023, \mn@doi
  [\apj] {10.3847/1538-4357/acbd3a}, \href
  {https://ui.adsabs.harvard.edu/abs/2023ApJ...946....3K} {946, 3}

\bibitem[\protect\citeauthoryear{{Kim}, {Armillotta}, {Ostriker}, {Moon},
  {Lancaster}, {Kim}, {Linzer}  \& {Hix}}{{Kim}
  et~al.}{2026}]{KimTIGRESSpp2026}
{Kim} C.-G.,  {Armillotta} L.,  {Ostriker} E.~C.,  {Moon} S.,  {Lancaster} L.,
  {Kim} J.-G.,  {Linzer} N.~B.,   {Hix} R.~N.,  2026, \mn@doi [\apj]
  {10.3847/1538-4357/ae76d6}, \href
  {https://ui.adsabs.harvard.edu/abs/2026ApJ..1006..123K} {1006, 123}

\bibitem[\protect\citeauthoryear{{Klein}, {McKee}  \& {Colella}}{{Klein}
  et~al.}{1994}]{Klein1994}
{Klein} R.~I.,  {McKee} C.~F.,   {Colella} P.,  1994, \mn@doi [\apj]
  {10.1086/173554}, \href
  {https://ui.adsabs.harvard.edu/abs/1994ApJ...420..213K} {420, 213}

\bibitem[\protect\citeauthoryear{{Koyama} \& {Inutsuka}}{{Koyama} \&
  {Inutsuka}}{2002}]{Koyama2002}
{Koyama} H.,  {Inutsuka} S.-i.,  2002, \mn@doi [\apjl] {10.1086/338978}, \href
  {https://ui.adsabs.harvard.edu/abs/2002ApJ...564L..97K} {564, L97}

\bibitem[\protect\citeauthoryear{{Krieger} et~al.,}{{Krieger}
  et~al.}{2019}]{Krieger19a}
{Krieger} N.,  et~al., 2019, \mn@doi [\apj] {10.3847/1538-4357/ab2d9c}, \href
  {https://ui.adsabs.harvard.edu/abs/2019ApJ...881...43K} {881, 43}

\bibitem[\protect\citeauthoryear{{Krumholz}, {Fumagalli}, {da Silva}, {Rendahl}
   \& {Parra}}{{Krumholz} et~al.}{2015}]{Krumholz15b}
{Krumholz} M.~R.,  {Fumagalli} M.,  {da Silva} R.~L.,  {Rendahl} T.,   {Parra}
  J.,  2015, \mn@doi [\mnras] {10.1093/mnras/stv1374}, \href
  {http://adsabs.harvard.edu/abs/2015MNRAS.452.1447K} {452, 1447}

\bibitem[\protect\citeauthoryear{{Leidi}, {Andrassy}, {Barsukow}, {Higl},
  {Edelmann}  \& {R{\"o}pke}}{{Leidi} et~al.}{2024}]{Leidi2024}
{Leidi} G.,  {Andrassy} R.,  {Barsukow} W.,  {Higl} J.,  {Edelmann} P.~V.~F.,
  {R{\"o}pke} F.~K.,  2024, \mn@doi [\aap] {10.1051/0004-6361/202348882}, \href
  {https://ui.adsabs.harvard.edu/abs/2024A&A...686A..34L} {686, A34}

\bibitem[\protect\citeauthoryear{{Leitherer} et~al.,}{{Leitherer}
  et~al.}{1999}]{Leitherer99a}
{Leitherer} C.,  et~al., 1999, \mn@doi [\apjs] {10.1086/313233}, \href
  {http://adsabs.harvard.edu/cgi-bin/nph-bib\_query?bibcode=1999ApJS..123....3L\&db\_key=AST}
  {123, 3}

\bibitem[\protect\citeauthoryear{{Leroy} et~al.,}{{Leroy}
  et~al.}{2015}]{Leroy15b}
{Leroy} A.~K.,  et~al., 2015, \mn@doi [\apj] {10.1088/0004-637X/814/2/83},
  \href {http://adsabs.harvard.edu/abs/2015ApJ...814...83L} {814, 83}

\bibitem[\protect\citeauthoryear{{Lockman}, {Di Teodoro}  \&
  {McClure-Griffiths}}{{Lockman} et~al.}{2020}]{Lockman2020}
{Lockman} F.~J.,  {Di Teodoro} E.~M.,   {McClure-Griffiths} N.~M.,  2020,
  \mn@doi [\apj] {10.3847/1538-4357/ab55d8}, \href
  {https://ui.adsabs.harvard.edu/abs/2020ApJ...888...51L} {888, 51}

\bibitem[\protect\citeauthoryear{{Martini}, {Leroy}, {Mangum}, {Bolatto},
  {Keating}, {Sandstrom}  \& {Walter}}{{Martini} et~al.}{2018}]{Martini18a}
{Martini} P.,  {Leroy} A.~K.,  {Mangum} J.~G.,  {Bolatto} A.,  {Keating} K.~M.,
   {Sandstrom} K.,   {Walter} F.,  2018, \mn@doi [\apj]
  {10.3847/1538-4357/aab08e}, \href
  {https://ui.adsabs.harvard.edu/abs/2018ApJ...856...61M} {856, 61}

\bibitem[\protect\citeauthoryear{{McClure-Griffiths}, {Green}, {Hill},
  {Lockman}, {Dickey}, {Gaensler}  \& {Green}}{{McClure-Griffiths}
  et~al.}{2013}]{McClureGriffiths2013}
{McClure-Griffiths} N.~M.,  {Green} J.~A.,  {Hill} A.~S.,  {Lockman} F.~J.,
  {Dickey} J.~M.,  {Gaensler} B.~M.,   {Green} A.~J.,  2013, \mn@doi [\apjl]
  {10.1088/2041-8205/770/1/L4}, \href
  {https://ui.adsabs.harvard.edu/abs/2013ApJ...770L...4M} {770, L4}

\bibitem[\protect\citeauthoryear{{McCourt}, {O'Leary}, {Madigan}  \&
  {Quataert}}{{McCourt} et~al.}{2015}]{McCourt2015}
{McCourt} M.,  {O'Leary} R.~M.,  {Madigan} A.-M.,   {Quataert} E.,  2015,
  \mn@doi [\mnras] {10.1093/mnras/stv355}, \href
  {https://ui.adsabs.harvard.edu/abs/2015MNRAS.449....2M} {449, 2}

\bibitem[\protect\citeauthoryear{{Noon}, {Krumholz}, {Di Teodoro},
  {McClure-Griffiths}, {Lockman}  \& {Armillotta}}{{Noon}
  et~al.}{2023}]{Noon2023}
{Noon} K.~A.,  {Krumholz} M.~R.,  {Di Teodoro} E.~M.,  {McClure-Griffiths}
  N.~M.,  {Lockman} F.~J.,   {Armillotta} L.,  2023, \mn@doi [\mnras]
  {10.1093/mnras/stad1890}, \href
  {https://ui.adsabs.harvard.edu/abs/2023MNRAS.524.1258N} {524, 1258}

\bibitem[\protect\citeauthoryear{{Noon}, {Krumholz}, {McClure-Griffiths}, {Di
  Teodoro}  \& {Armillotta}}{{Noon} et~al.}{2026}]{Noon26a}
{Noon} K.~A.,  {Krumholz} M.~R.,  {McClure-Griffiths} N.~M.,  {Di Teodoro}
  E.~M.,   {Armillotta} L.,  2026, \mn@doi [\mnras] {10.1093/mnras/stag715},
  \href {https://ui.adsabs.harvard.edu/abs/2026MNRAS.548ag715N} {548, stag715}

\bibitem[\protect\citeauthoryear{{Peschken}, {Di Teodoro}, {Hanasz},
  {Armillotta}, {Tress}, {Naab}, {W{\'o}lta{\'n}ski}  \&
  {Gawryszczak}}{{Peschken} et~al.}{2026}]{Peschken2026}
{Peschken} N.,  {Di Teodoro} E.~M.,  {Hanasz} M.,  {Armillotta} L.,  {Tress}
  R.,  {Naab} T.,  {W{\'o}lta{\'n}ski} D.,   {Gawryszczak} A.,  2026, \mn@doi
  [\aap] {10.1051/0004-6361/202659639}, \href
  {https://ui.adsabs.harvard.edu/abs/2026A&A...712A.148P} {712, A148}

\bibitem[\protect\citeauthoryear{{Rathjen}, {Naab}, {Walch}, {Seifried},
  {Girichidis}  \& {W{\"u}nsch}}{{Rathjen} et~al.}{2023}]{Rathjen2023}
{Rathjen} T.-E.,  {Naab} T.,  {Walch} S.,  {Seifried} D.,  {Girichidis} P.,
  {W{\"u}nsch} R.,  2023, \mn@doi [\mnras] {10.1093/mnras/stad1104}, \href
  {https://ui.adsabs.harvard.edu/abs/2023MNRAS.522.1843R} {522, 1843}

\bibitem[\protect\citeauthoryear{{Schneider} \& {Mao}}{{Schneider} \&
  {Mao}}{2024}]{SchneiderMao2024}
{Schneider} E.~E.,  {Mao} S.~A.,  2024, \mn@doi [\apj]
  {10.3847/1538-4357/ad2e8a}, \href
  {https://ui.adsabs.harvard.edu/abs/2024ApJ...966...37S} {966, 37}

\bibitem[\protect\citeauthoryear{{Schneider} \& {Robertson}}{{Schneider} \&
  {Robertson}}{2017}]{SchneiderRobertson2017}
{Schneider} E.~E.,  {Robertson} B.~E.,  2017, \mn@doi [\apj]
  {10.3847/1538-4357/834/2/144}, \href
  {https://ui.adsabs.harvard.edu/abs/2017ApJ...834..144S} {834, 144}

\bibitem[\protect\citeauthoryear{{Schneider} \& {Robertson}}{{Schneider} \&
  {Robertson}}{2018}]{SchneiderRobertson2018}
{Schneider} E.~E.,  {Robertson} B.~E.,  2018, \mn@doi [\apj]
  {10.3847/1538-4357/aac329}, \href
  {https://ui.adsabs.harvard.edu/abs/2018ApJ...860..135S} {860, 135}

\bibitem[\protect\citeauthoryear{{Schneider}, {Robertson}  \&
  {Thompson}}{{Schneider} et~al.}{2018}]{Schneider2018}
{Schneider} E.~E.,  {Robertson} B.~E.,   {Thompson} T.~A.,  2018, \mn@doi
  [\apj] {10.3847/1538-4357/aacce1}, \href
  {https://ui.adsabs.harvard.edu/abs/2018ApJ...862...56S} {862, 56}

\bibitem[\protect\citeauthoryear{{Schneider}, {Ostriker}, {Robertson}  \&
  {Thompson}}{{Schneider} et~al.}{2020}]{Schneider2020}
{Schneider} E.~E.,  {Ostriker} E.~C.,  {Robertson} B.~E.,   {Thompson} T.~A.,
  2020, \mn@doi [\apj] {10.3847/1538-4357/ab8ae8}, \href
  {https://ui.adsabs.harvard.edu/abs/2020ApJ...895...43S} {895, 43}

\bibitem[\protect\citeauthoryear{{Shah}, {Krumholz}  \&
  {McClure-Griffiths}}{{Shah} et~al.}{2025}]{Shah25a}
{Shah} H.,  {Krumholz} M.~R.,   {McClure-Griffiths} N.~M.,  2025, \mn@doi
  [\mnras] {10.1093/mnras/staf1003}, \href
  {https://ui.adsabs.harvard.edu/abs/2025MNRAS.541.1106S} {541, 1106}

\bibitem[\protect\citeauthoryear{{Sike}, {Thomas}, {Ruszkowski}, {Pfrommer}  \&
  {Weber}}{{Sike} et~al.}{2025}]{Sike25a}
{Sike} B.,  {Thomas} T.,  {Ruszkowski} M.,  {Pfrommer} C.,   {Weber} M.,  2025,
  \mn@doi [\apj] {10.3847/1538-4357/adda3d}, \href
  {https://ui.adsabs.harvard.edu/abs/2025ApJ...987..204S} {987, 204}

\bibitem[\protect\citeauthoryear{{Smith} et~al.,}{{Smith}
  et~al.}{2017}]{Smith2017}
{Smith} B.~D.,  et~al., 2017, \mn@doi [\mnras] {10.1093/mnras/stw3291}, \href
  {https://ui.adsabs.harvard.edu/abs/2017MNRAS.466.2217S} {466, 2217}

\bibitem[\protect\citeauthoryear{{Smith} et~al.,}{{Smith}
  et~al.}{2024}]{Smith24a}
{Smith} M.~C.,  et~al., 2024, \mn@doi [\mnras] {10.1093/mnras/stad3168}, \href
  {https://ui.adsabs.harvard.edu/abs/2024MNRAS.527.1216S} {527, 1216}

\bibitem[\protect\citeauthoryear{{Sukhbold}, {Ertl}, {Woosley}, {Brown}  \&
  {Janka}}{{Sukhbold} et~al.}{2016}]{Sukhbold16a}
{Sukhbold} T.,  {Ertl} T.,  {Woosley} S.~E.,  {Brown} J.~M.,   {Janka} H.-T.,
  2016, \mn@doi [\apj] {10.3847/0004-637X/821/1/38}, \href
  {http://adsabs.harvard.edu/abs/2016ApJ...821...38S} {821, 38}

\bibitem[\protect\citeauthoryear{{Tan} \& {Fielding}}{{Tan} \&
  {Fielding}}{2024}]{TanFielding2024}
{Tan} B.,  {Fielding} D.~B.,  2024, \mn@doi [\mnras] {10.1093/mnras/stad3793},
  \href {https://ui.adsabs.harvard.edu/abs/2024MNRAS.527.9683T} {527, 9683}

\bibitem[\protect\citeauthoryear{{Thomas}, {Pfrommer}  \& {Pakmor}}{{Thomas}
  et~al.}{2025}]{Thomas25a}
{Thomas} T.,  {Pfrommer} C.,   {Pakmor} R.,  2025, \mn@doi [\aap]
  {10.1051/0004-6361/202450817}, \href
  {https://ui.adsabs.harvard.edu/abs/2025A&A...698A.104T} {698, A104}

\bibitem[\protect\citeauthoryear{{V{\'a}zquez} \& {Leitherer}}{{V{\'a}zquez} \&
  {Leitherer}}{2005}]{Vazquez05a}
{V{\'a}zquez} G.~A.,  {Leitherer} C.,  2005, \mn@doi [\apj] {10.1086/427866},
  \href
  {http://adsabs.harvard.edu/cgi-bin/nph-bib\_query?bibcode=2005ApJ...621..695V\&db\_key=AST}
  {621, 695}

\bibitem[\protect\citeauthoryear{{Veilleux}, {Maiolino}, {Bolatto}  \&
  {Aalto}}{{Veilleux} et~al.}{2020}]{Veilleux20a}
{Veilleux} S.,  {Maiolino} R.,  {Bolatto} A.~D.,   {Aalto} S.,  2020, \mn@doi
  [\aapr] {10.1007/s00159-019-0121-9}, \href
  {https://ui.adsabs.harvard.edu/abs/2020A&ARv..28....2V} {28, 2}

\bibitem[\protect\citeauthoryear{Vijayan, Krumholz  \& Wibking}{Vijayan
  et~al.}{2024}]{QEDI}
Vijayan A.,  Krumholz M.~R.,   Wibking B.~D.,  2024, \mn@doi [\mnras]
  {10.1093/mnras/stad3816}, 527, 10095

\bibitem[\protect\citeauthoryear{Vijayan, Krumholz  \& Wibking}{Vijayan
  et~al.}{2025}]{QEDIII}
Vijayan A.,  Krumholz M.~R.,   Wibking B.~D.,  2025, \mn@doi [\mnras]
  {10.1093/mnras/staf136}, 539, 1706

\bibitem[\protect\citeauthoryear{{Vijayan}, {Krumholz}  \& {Wibking}}{{Vijayan}
  et~al.}{2026}]{QEDV}
{Vijayan} A.,  {Krumholz} M.~R.,   {Wibking} B.~D.,  2026, \mn@doi [\mnras]
  {10.1093/mnras/stag913}, \href
  {https://ui.adsabs.harvard.edu/abs/2026MNRAS.549ag913V} {549, stag913}

\bibitem[\protect\citeauthoryear{{Walch} et~al.,}{{Walch}
  et~al.}{2015}]{Walch2015}
{Walch} S.,  et~al., 2015, \mn@doi [\mnras] {10.1093/mnras/stv1975}, \href
  {https://ui.adsabs.harvard.edu/abs/2015MNRAS.454..238W} {454, 238}

\bibitem[\protect\citeauthoryear{Wibking \& Krumholz}{Wibking \&
  Krumholz}{2022}]{Wibking2022}
Wibking B.~D.,  Krumholz M.~R.,  2022, \mn@doi [\mnras]
  {10.1093/mnras/stac439}, 512, 1430

\bibitem[\protect\citeauthoryear{{Yuan}, {Krumholz}  \& {Martin}}{{Yuan}
  et~al.}{2023}]{Yuan23a}
{Yuan} Y.,  {Krumholz} M.~R.,   {Martin} C.~L.,  2023, \mn@doi [\mnras]
  {10.1093/mnras/stac3241}, \href
  {https://ui.adsabs.harvard.edu/abs/2023MNRAS.518.4084Y} {518, 4084}

\bibitem[\protect\citeauthoryear{{da Silva}, {Fumagalli}  \& {Krumholz}}{{da
  Silva} et~al.}{2012}]{da-Silva12a}
{da Silva} R.~L.,  {Fumagalli} M.,   {Krumholz} M.,  2012, \mn@doi [\apj]
  {10.1088/0004-637X/745/2/145}, \href
  {http://adsabs.harvard.edu/abs/2012ApJ...745..145D} {745, 145}

\makeatother
\end{thebibliography}

\appendix

\section{The cooling table}
\label{sec:table}

\begin{figure}
\centering
\includegraphics[width=\columnwidth]{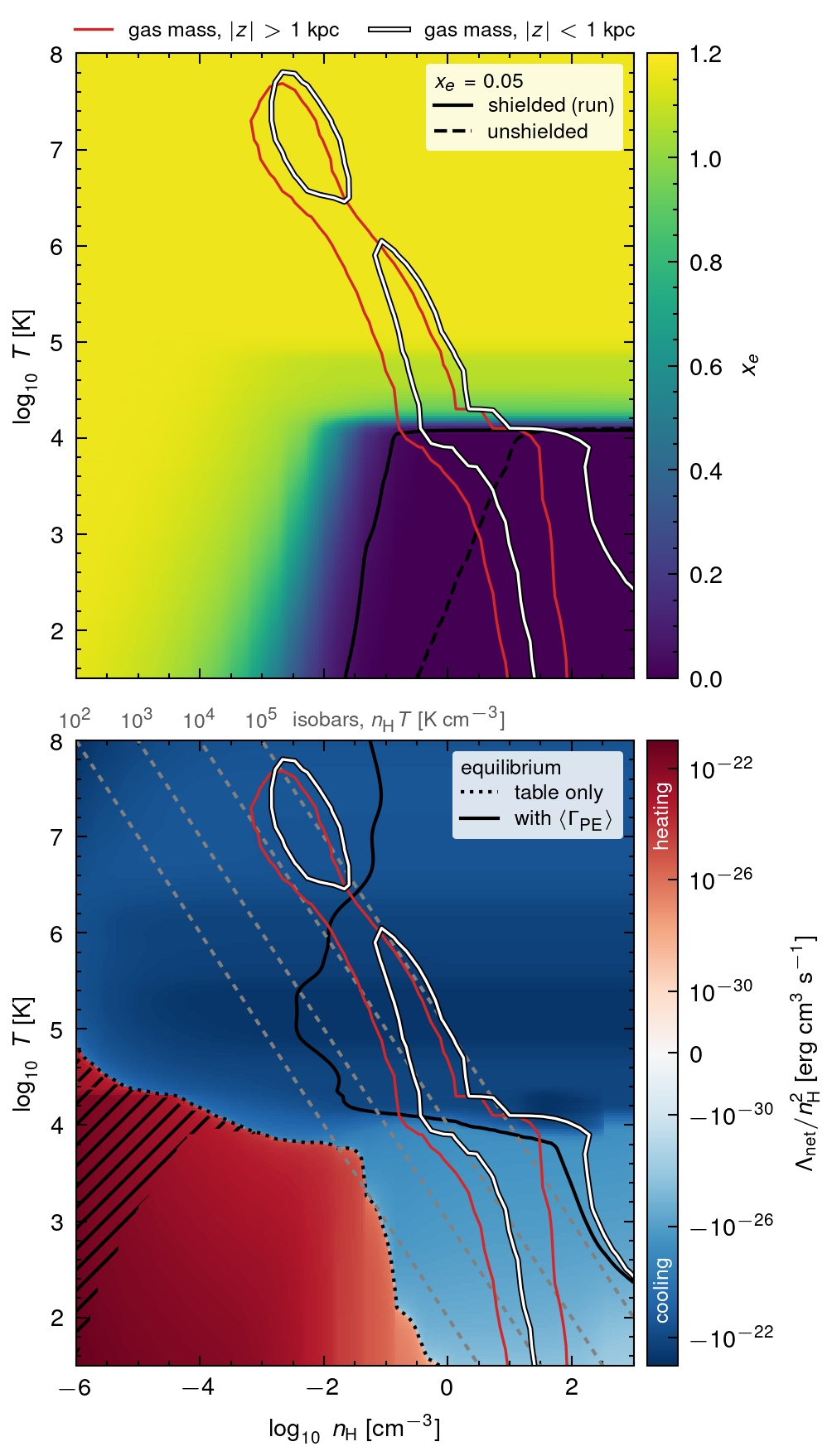}
\caption{Comparison of simulations results with cooling tables. Top: color shows the electron abundance $x_e$ of the shielded table used in the simulations as a function of gas number density $n_{\rm H}$ and temperature $T$. The solid black line shows the boundary $x_e = 0.05$ we use to define neutral material. For comparison, the dashed black line shows the boundary for an unshielded table. Bottom: colors show net heating (positive, red) or cooling (negative, blue) rate per $n_{\rm H}^2$, $\Lambda_{\rm net}/n_{\rm H}^2$, of the shielded table without the photoelectric term, on a symmetric logarithmic scale that is linear for $|\Lambda_{\rm net}|/n_{\rm H}^2 < 10^{-30}$~erg~cm$^3$~s$^{-1}$. The dotted black line shows the equilibrium temperature with no photoelectric heating, and the solid black line shows the equilibrium at the mean photoelectric heating rate of the $1$~pc run over $29$--$112$~Myr. Grey dashed lines are isobars, labelled by $n_{\rm H} T$ in K~cm$^{-3}$. The hatched region is where the shielded table heats more than twice as fast as the unshielded one, the excess heating at low density noted in the \textsc{Grackle} documentation. In both upper and lower panels, red and white contours enclose the region that contains $99$ per cent of the gas mass above and below $|z| = 1$~kpc in the $1$~pc run, summed over the snapshots at $36$, $40$, $52$ and $56$~Myr.}
\label{fig:table}
\end{figure}

The most uncertain part of the choice of cooling table is the shielding prescription, because no single prescription provides a good description of gas both in the disc and in the circumgalactic medium. Our chosen \textsc{Grackle} table attenuates the \citet{Haardt2012} background by a column equal to the gas density times the local Jeans length, capped at $100$~pc \citep{Smith2017}, and while this prescription is reasonable in the disc, it overestimates the shielding in atomic gas in the wind that is not organised into self-gravitating structures: at the densities of our neutral wind clouds this column is up to eight times larger than the clouds' measured columns (Table~\ref{tab:clouds}). This conceivably causes us to overestimate the amount of neutral material in the wind. This difficulty is the reason that simulations that seek to reproduce the detailed ionisation structure of the gas in both the disc and the CGM \citep[e.g.,][]{Shah25a} generally switch between tables with different shielding prescriptions.

To quantify by how much our choice of table could be causing us to overestimate the neutral mass in the wind, we repeat our analysis of the 1 pc and 2 pc resolution runs using the \textsc{Grackle} table \texttt{CloudyData\_UVB=HM2012.h5}, which uses the same radiation field but assumes no attenuation. We emphasise that this is a deliberately extreme choice: while our fiducial table overestimates the shielding column in our neutral clouds, our measured columns, $N_{\rm H} \approx 5\times10^{19}$--$10^{21}$~cm$^{-2}$, exceed by more than two orders of magnitude the column of $\sim 10^{17}$~cm$^{-2}$ at which atomic hydrogen becomes optically thick to ionizing photons near the ionization threshold. Thus by using this table we set a very conservative upper limit on the effects of choosing a different shielding prescription.

Carrying out this exercise, we find that using the unshielded table lowers $\eta_{\rm atom}$ by only $5$--$15$ per cent and $13$--$20$ per cent in the 1 pc and 2 pc runs, respectively. The atomic content of the wind as it would be measured by a 21 cm observation is therefore robust to the treatment of shielding. However, using the unshielded table does significantly reduce the amount of material with electron abundance $x_e < 0.05$, because it raises the electron fraction in predominantly neutral gas at moderate densities and temperatures from $x_e \sim 0.01$ to $x_e \sim 0.1$. Such gas is still predominantly neutral, which is why $\eta_{\rm atom}$ changes little, but it no longer qualifies as neutral by the strict conditions we use in the main text, and thus $\eta_{\him}$ as defined by the threshold drops by a factor of about $5$--$20$ at $\Delta x = 1$~pc and $20$--$50$ at $\Delta x = 2$~pc, depending on height.

We visualise this effect in the upper panel of Figure~\ref{fig:table}, which shows the electron abundance of the shielded \textsc{Grackle} table used in our simulations as functions of hydrogen density and temperature, together with the mass distribution of the gas in the $1$~pc run (red and white contours, showing the gas at $|z| > 1$ kpc and $<1$ kpc, respectively). The solid black contour shows the locus $x_e = 0.05$ in the shielded table, and for comparison the dashed contour shows this locus in the unshielded table. We see that using the unshielded table effectively reclassifies a large amount of gas at $n_\mathrm{H} \sim 1$ cm$^{-3}$ and $T\sim 5000$ K from neutral to WIM. However, we emphasise that this is a classification effect only, and that if we were simply to shift our threshold for ``neutral cloud'' from $x_e < 0.05$ to $x_e < 0.2$, for example, the change in $\eta_{\him}$ would disappear.

Moreover, we expect the choice of cooling table to have negligible effects on the \textit{dynamics}
of the predominantly neutral clouds. Condensation onto the clouds is controlled by cooling in the mixing layers: at the mixing-layer temperature $T_{\rm mix} \approx 2\times10^5$~K the cooling rates of the two tables differ by $2$ per cent, and the time for mixed gas to cool at constant pressure from $2\times10^6$~K to $1.5\times10^4$~K differs by $10$ per cent; the tables differ by up to a factor of two only just above $10^4$~K, where the gas has almost finished cooling. In both tables the equilibrium temperature of the cloud gas is close to $10^4$~K. We therefore conclude that, while the amount of gas that is collected into almost entirely neutral clouds ($x_e < 0.05$) is sensitive to the choice of cooling table, neither the total atomic mass flux in the outflow nor the dynamics are.

The table we use also has a second known limitation: the \textsc{Grackle} documentation warns that the shielded tables have unphysically high heating rates at low densities and does not recommend them in the tabulated mode that we use. We identify this region in the lower panel of Figure~\ref{fig:table}; in this figure, the hatched black region is the locus where the heating rate of the shielded table is more than twice the heating rate of the unshielded one. It is clear that the region where this concern arises is limited to $n_{\rm H} \lesssim 10^{-3}$~cm$^{-3}$ and $T \lesssim 10^{4.8}$~K, and that this locus contains negligible gas. The reason this region is unpopulated is that it corresponds to pressures far lower than will be found in a galactic disc simulation such as ours; the dashed grey lines show isobars from $P/k_{\rm B} = 10^2 - 10^6$ K cm$^{-3}$, straddling a range around the gravitational pressure supplied by the stellar disc, $P/k_{\rm B} \approx G \Sigma_{\rm gas} \Sigma_*/k_{\rm B} \approx 2\times 10^5$ K cm$^{-3}$ for the initial gas and stellar surface densities in our simulations. We see that even gas at a pressure $\approx 3$ orders of magnitude below the characteristic pressure in our simulation domain still remains several orders of magnitude above the range of pressures where excessive heating is a concern. We conclude that the overheating problem is relevant only for simulations that reach outer CGM-like or even IGM-like conditions, and is not a concern for galaxy simulations such as ours.

\section{Lagrangian accelerations}
\label{sec:lagrangian}

We estimate the Lagrangian vertical acceleration of the gas in a given phase as a function of height off the plane $z$,
\begin{equation}
    \frac{Dv_z}{Dt} = \left.\frac{\partial v_z}{\partial t}\right|_z + v_z\left.\frac{\partial v_z}{\partial z}\right|_t,
\end{equation}
as follows. The advective term $v_z(\partial v_z/\partial z)_t$ is a spatial derivative at one instant, which we evaluate from the single snapshot at $t = 60$~Myr. For this snapshot, we first compute the mass-weighted mean velocity of the gas in the phase of interest at height $z$ by averaging in slabs $\Delta z = 0.6$ kpc thick, and averaging over the two sides of the midplane; we apply this slab-averaging procedure to reduce small-scale noise, which would otherwise dominate our estimate of the derivative. We then compute the derivative of the vertical velocity with respect to $z$ using centred differences on these slab averages. To estimate the Eulerian term $(\partial v_z/\partial t)_z$, we compute the same $\Delta z = 0.6$ kpc-slab averages on the $t=55.4$ and 64 Myr snapshots, and estimate $(\partial v_z/\partial t)_z$ at $t=60$ Myr from a centred difference between these two times.

We obtain estimates of the uncertainty on the spatial derivative at fixed time by varying the slab width from $0.5$ to $1.5$ times our fiducial value, and on the time derivative by computing two one-sided differences from the same three time slices (i.e., from 55.4 to 60 Myr and from 60 to 64 Myr) and taking half the difference between the two one-sided differences as our uncertainty estimate. We obtain a total uncertainty on the final Lagrangian derivative $Dv_z/Dt$ by adding the spatial and temporal uncertainties in quadrature.

\section{Calculation of the acceleration by ram pressure and condensation}
\label{sec:drag}

Evaluating the terms describing gas acceleration by ram pressure / drag and condensation in equation~(\ref{eq:budget}) (Section~\ref{sec:accel}) requires that we compute or adopt estimates for the cloud mass $M$, cross-sectional area $A$, the density of the hot gas incident on the cloud $\rho_\mathrm{h}$, the hot gas-H~\textsc{i} velocity difference $\Delta v$, the condensation rate $\Gamma_\mathrm{h\to\him}$, and coefficient of drag $C_D$. We estimate these terms using the sample population of neutral clouds produced by our segmentation procedure, as described in Section~\ref{sec:fullres}, which provides a catalogue of H~\textsc{i} clouds at five times from $49-70.6$ Myr and at three heights, centred on $1.5$, $2.5$, and $3.5$ kpc.

We have already defined the mass $M$ for each of these clouds (Section~\ref{sec:fullres}). To compute the cross-sectional area $A$, we flatten the set of cells defining the cloud onto the $xy$ plane and computing the area of the resulting set of pixels; mathematically, our definition is
\begin{equation}
    A = \Delta x^2 \sum_{ij} \max_k(I_{\mathrm{cloud},ijk})
\end{equation}
where $I_{\mathrm{cloud},ijk}$ is an indicator function on cell $ijk$ which is unity if a cell is identified as belonging to a cloud and 0 otherwise. We take the ratio $M/A$ that appears in the drag term in equation~(\ref{eq:budget}) to be simply the mean cloud mass divided by the mean cloud area in each slab.

We consider two possible estimates for the hot gas density $\rho_\mathrm{h}$ and hot-neutral velocity difference $\Delta v$, one based on mean properties and one based on properties in cloud boundary layers. For our mean property-based estimate, we simply evaluate $\rho_\mathrm{h}$ and $\Delta v = v_{\mathrm{h},z} - v_{{\him},z}$ using the slab-averaged means shown in Fig.~\ref{fig:zprof} at $t = 60$ Myr, and from the analogous quantities computed at our other sample times. For our boundary-layer estimate, we use the same value of $v_{{\him},z}$ for the mean H~\textsc{i} velocity, and we estimate mean values of the hot gas density and velocity as follows:
\begin{enumerate}
    \item Define an initial boundary shell around each H~\textsc{i} cloud by starting with the cells of the cloud itself and applying $N_d = \max[3, \operatorname{round}(R_\mathrm{eff}/\Delta x)]$ iterations of binary dilation with a structuring element with a connectivity of 1. This defines a shell around each cloud with a characteristic radius comparable to the cloud size, but with a minimum size of three cells, and whose shape is contoured to the cloud shape.
    \item Next, remove any neutral or non-hot cells, those with $x_e < 0.05$ or $T < 10^6$ K, leaving only cells classified as hot in our phase decomposition.
    \item Finally, discard any cells for which the distance from the midplane to their centre, $|z|$, is larger than the cloud's distance to the midplane, defined as the median $|z|$ of cells that belong to the cloud. This step means that we keep only the parts of the shell that are closer to the midplane than cloud centre, and are therefore upstream of it.
\end{enumerate}
Once we have defined the cells belonging to the boundary layer around each cloud, we compute the mean density $\rho_\mathrm{h}$ and mass-weighted mean $z$ velocity $v_{\mathrm{h},z}$ over all the cells tagged as belonging to a boundary layer in each slab, and use these for our boundary-layer estimate of the drag / ram pressure force.

To estimate the condensation rate $\Gamma_{\rm h\to\him}$ in each of our sample slabs, we note that, for any phase $p$, the rate of change in the total mass $M_p$ of that phase within a slab from $z_0$ to $z_1$ obeys
\begin{equation}
\frac{\mathrm{d}M_p}{\mathrm{d}t}
   = \dot{M}_p(z_0,t) - \dot{M}_p(z_1,t) + \Gamma_p ,
\label{eq:source}
\end{equation}
where $\dot{M}_p(h,t)$ is the outward mass fluxes through the surface at height $h$ (equation~\ref{eq:mass_flux}) and $\Gamma_p$ is the net rate at which phase $p$ gains mass from the other two. We estimate $\mathrm{d}M_p/\mathrm{d}t$ for each phase over our five sample snapshots by computing the mass $M_p$ of each phase inside each slab at each time, and then carrying out a simple least-squares fit to find the slope. We directly measure the fluxes from the simulation at each time; using these fluxes together with our estimate of $\mathrm{d}M_p/\mathrm{d}t$ yields an estimate for $\Gamma_p$ at each time, which is the quantity of interest. Rather than use $\Gamma_{\him}$ estimated in this way directly, we instead evaluate it as $\Gamma_{\him} = -(\Gamma_{\rm hot} + \Gamma_{\rm WIM})$; we do so because the time series of H~\textsc{i} mass and instantaneous mass fluxes are rather noisy due to the clumpy H~\textsc{i} structure. The WIM and hot gas masses vary much more smoothly in time, allowing significantly better estimates.

Finally, for $C_D$ plausible values are in the range $C_D \approx 1 - 2$, with smaller values typically characterising smoothly tapered structures such as spheres and larger values for smooth or blunt bodies. We compute estimates for both $C_D = 1$ and 2 to give a plausible uncertainty range.

\bsp	%
\label{lastpage}
\end{document}